\documentclass[reprint,groupedaddress,superscriptaddress,amsfonts,showkeys]{revtex4-1}
\usepackage[T1]{fontenc}
\usepackage{amsmath,amssymb,dcolumn,bm,pifont}
\usepackage{siunitx}
\usepackage[pdftex]{graphicx}
\usepackage[caption=false]{subfig}
\usepackage{xcolor}
\DeclareGraphicsExtensions{{.pdf},{.png}}
\graphicspath{{figures/png/}}
\newcommand{\new}[1]{\textcolor{red}{#1}}

\newtheorem{example}{Example}
\def\sinc{\mathop{\rm sinc}}

\def\RF{\textsc{RF}}
\def\SE{\textsc{SE}}
\def\UDD{\textsc{UDD}}
\def\CPMG{\textsc{CPMG}}
\def\FSE{\textsc{FSE}}
\def\SINC{\textsc{SINC}}

\begin{document}
\title[Contrast enhancement by novel dynamic decoupling sequences]
      {Comparison of CPMG and Uhrig Dynamic Decoupling (UDD) for tissue refocusing in MRI}
      
\author{Sophie Shermer}
\address{Department of Physics, College of Science, Swansea University,  Singleton Park, Swansea SA2 8PP, United Kingdom}
\author{Jonathan Phillips}
\address{Institute of Life Science, Medical School, Swansea University, Singleton Park, Swansea SA2 8PP, United Kingdom}

\date{\today}


\begin{abstract}
Two different dynamic decoupling strategies, the classic \CPMG\ and the more recently proposed \UDD\ protocol are compared in the context of magnetic resonance imaging (MRI).   Both sequences were implemented on a 3T human MRI system and relaxometry was performed for a variety of tissue-mimicking agarose gels.  We find that \CPMG\ provides moderately better decoupling than \UDD.  The results are consistent with experimental noise spectra obtained from \textit{in vivo} spectroscopy results, which suggest that the type of noise spectra obtained would favor \CPMG.  Theoretical coherence pathway analysis further suggests that \UDD\ is more susceptible to systematic errors due to static magnetic and radio-frequency field inhomogeneity or imperfect excitation and refocusing pulse profiles.
\end{abstract}


\maketitle

\section{Introduction}

Refocusing is a technique widely used in nuclear magnetic resonance (NMR) to compensate for systematic errors, \textit{e.g.}, arising from static magnetic field inhomogeneities, and as a way to modify image contrast in MRI.  The oldest technique is the Hahn spin echo (\SE), which involves a $180^\circ$ refocusing pulse inserted halfway between the excitation and readout pulse~\cite{Hahn1950}.  Modern fast spin echo (\FSE) techniques utilize the same basic 90$^\circ$-180$^\circ$ \SE\ but with multiple 180$^\circ$ refocusing pulses to reduce image acquisition times.  \SE\ and \FSE\ sequences form the basis of many clinical imaging protocols.

Hahn's work was later extended by Carr, Purcell, Meiboom and Gill~\cite{Carr1954,Meiboom1958}.  The resulting \CPMG\ sequence consists of two $180^\circ$ refocusing pulses applied at time $\tfrac{1}{4}T_E$ and $\tfrac{3}{4}T_E$, where $T_E$ is the echo readout time.  Unlike for the Hahn spin echo, the refocusing pulses are usually applied about an axis orthogonal to that of the excitation pulse.  The two-pulse \CPMG\ sequence can be repeated $n$ times, resulting in a sequence of $2n$ regularly spaced refocusing pulses.  Regular spacing of the refocusing pulses was assumed to be optimal, especially in mitigating the effects of free diffusion~\cite{Meiboom1958}.

More recently Uhrig showed that changing the spacings between the $n$ refocusing pulses can improve decoupling for a spin-$\frac{1}{2}$ particle coupled to a noisy environment characterized by an Ohmic bath~\cite{Uhrig2007}.  In \UDD\ the pulse spacings are chosen to make the first $n-1$ derivatives of the modulation function with respect to the frequency $\omega$ vanish at $\omega=0$, the modulation function being the Fourier transform of the piecewise constant switching function that changes sign whenever a decoupling or refocusing pulse is applied.  It was later argued that this choice of timings should provide better refocusing 
for a range of noise spectra, not limited to Ohmic-bath-like spectra~\cite{Yang2008,Uhrig2010}.

This has spurred a number of recent papers investigating the efficiency of different dynamic decoupling (DD) schemes.  In Ref.~\cite{Szwer2011} it was observed that both \UDD\ and \CPMG\ afforded comparable protection from the ambient noise environment with regard to extending the coherence time of a memory qubit consisting of a single trapped $^{43}$Ca$^{+}$ ion.  Ref.~\cite{Du2009} found that \UDD\ outperformed \CPMG\ in preserving electron spin coherence in irradiated malonic acid crystals at a wide range of temperatures, while \CPMG\ outperformed \UDD\ for a qubit consisting of a $^{13}$C nuclear spin coupled to a proton spin bath with a close-to-Gaussian spectral density distribution~\cite{Ajoy2011}.

One of the applications suggested for \UDD\ is contrast enhancement in MRI.  Warren \emph{et al.} compared the \UDD\ sequence to the standard \CPMG\ sequence and concluded that it appeared to result in improvements in contrast in cancerous tissue in mice~\cite{Warren2009}.  However, the observed effects were generally small and signal differences could be due various non-idealities in the experimental setup.  In more recent work on endogenous magnetic resonance contrast based on the localized composition of fat, it was further observed that \CPMG\ appeared more effective in refocusing oil while \UDD\ appeared better at refocusing fat tissue \cite{Warren2013}.

These preliminary results suggest that advanced dynamic decoupling sequences such as \UDD\ could have interesting applications in MRI.  To assess this possibility and gain a better understanding the performance of novel decoupling sequence and their potential for applications in MRI, we perform relaxometry experiments to determine the observed transverse relaxation times for a set of tissue-mimicking agarose and agar hydrogel phantoms in a standard \SI{3}{T} human MRI scanner.  Specifically we determine the $T_2$ relaxation times for $n$-pulse refocusing sequences with \CPMG\ and \UDD\ spacing, respectively.  We further compare the performance of both sequences for different types of noise spectra using simulations and apply coherence pathway analysis to explain apparent differences in contrast observed in terms of differences in the sensitivity to tissue inhomogeneities as a result of interference effects for different echo pathways.

\section{Dynamic Decoupling and MRI contrast}

The main sources of image contrast in conventional MRI are differences in the spin density $\rho$ and effective longitudinal and transverse relaxation rates, $R_1=1/T_1$ and $R_2=1/T_2$ respectively, of proton spins in different tissues.  By decoupling the system from its environment to various degrees, dynamic decoupling pulses can change the effective relaxation rates and thus image contrast.

Inhomogeneities in the magnetic field, whether due to imperfections in the static magnetic field ($\vec{B}_0$), dynamic magnetic field gradients applied, tissue susceptibility effects or other sources, give rise to variations in the Larmor frequencies of the spins within a voxel.  As a result the proton spins precess at different frequencies and the ensemble dephases over time.  Applying a $180^\circ$ refocusing pulse halfway between the excitation and readout, inverts \new{the phase of the spins}, causing the ensemble to rephase at the target time.  Thus a Hahn spin echo effectively compensates for the static or zero-frequency component of the combined resonance offset resulting from all sources $\Delta\omega(\vec{r},t)$.  If the precession frequencies $\Delta\omega(\vec{r},t)$ are not constant in time then a single refocusing pulse will refocus the spins only partially.  Using average Hamiltonian theory it can be shown that refocusing can be improved by increasing the number of refocusing pulses.

Intuitively, equidistant spacing of the refocusing pulses, as in the repeated \CPMG\ scheme, appears the most natural choice.  However, it is not necessarily optimal.  Specifically, it was show in Ref.~\cite{Uhrig2007} that choosing the refocusing pulse timings $t_j$ according to the formula
\begin{equation}
  \label{eq:UDD-timing}
  t_j = \sin^2(\pi j/(2n+2)) T_E, \quad j=1,2,\ldots, n
\end{equation}
improves decoupling for an Ohmic bath noise spectrum with a high-frequency cut-off, in that it makes not only the zero-frequency component $\tilde{y}(0)$ of the modulation function vanish but also its first $n-1$ derivatives $\tilde{y}^{(n)}(\omega)$ at $\omega=0$.  To be precise, $\tilde{y}(\omega)$ is the Fourier transform of the time-domain switching function $y(t)$ of the toggling frame Hamiltonian $\tilde{H}=y(t)\hbar\Delta\omega(\vec{r})\sigma_z$, where $\sigma_z$ is Pauli matrix
\begin{equation*}
  \sigma_z = \frac{1}{2}\begin{pmatrix} 0 & 1 \\ 1 & 0 \end{pmatrix}.
\end{equation*}
For ideal instantaneous $180^\circ$ pulses applied at times $t_j$, $y(t)$ is piecewise constant, $\pm 1$, and each pulse simply changes its sign.  Ref.~\cite{Yang2008} showed using perturbation theory that the choice of pulse timings \eqref{eq:UDD-timing} suppresses pure dephasing to $N$th order for small $T$ not just for the spin-boson model for which it was originally derived~\cite{Uhrig2007} but for a much larger class of bath models.  It was also observed in this work that this result still holds for more general modulation functions, provided the scaled modulation function contains $f_n(\theta)$ only odd harmonics of $\sin((n+1)\theta)$.

Recent work further suggests that the refocusing performance of \UDD\ may depend on the tissue type~\cite{Warren2009}, and that the conventional \CPMG\ sequence may provide superior refocusing for liquids such as free water and oil, while \UDD\ may perform better for certain biological tissues such as fat~\cite{Warren2013}. 

\section{Experiment Design and Results}

The \CPMG\ and \UDD\ pulse sequences were implemented using the Siemens integrated development environment for applications (\texttt{IDEA}) C++ framework provided by the manufacturer.  All experiments were performed on a Siemens (Erlangen, Germany) \SI{3}{\tesla} Magnetom Skyra system with tissue-mimicking gel phantoms.

\subsection{Phantom preparation}

Five \SI{20}{ml} agarose gel samples with agarose concentrations ranging from 1\% to 5\% agarose (\#A0169, Sigma-Aldrich, Dorset, UK) were prepared by dissolving agarose powder in deionized water heated to $80-90^\circ$C while mixing for $30-40$ minutes.  Solidification and polymerization occurred at room temperature overnight.  In addition to these test-tube size gels we also prepared a larger cylindrical agar hydrogel phantom consisting of a glass jar with six gel layers with varying concentrations of agar agar ranging from \SIrange{0.25}{1.25}{g} per \SI{60}{ml} of deionized water slightly doped with manganese chloride to reduce the $T_1$ relaxation times.  Agarose, agar agar and carageenan gels were chosen due to their tissue-mimicking properties~\cite{Mitchell1986,2016arXiv160808542G}.
  
\subsection{Sequence design}

\begin{figure*}
\includegraphics[width=\textwidth]{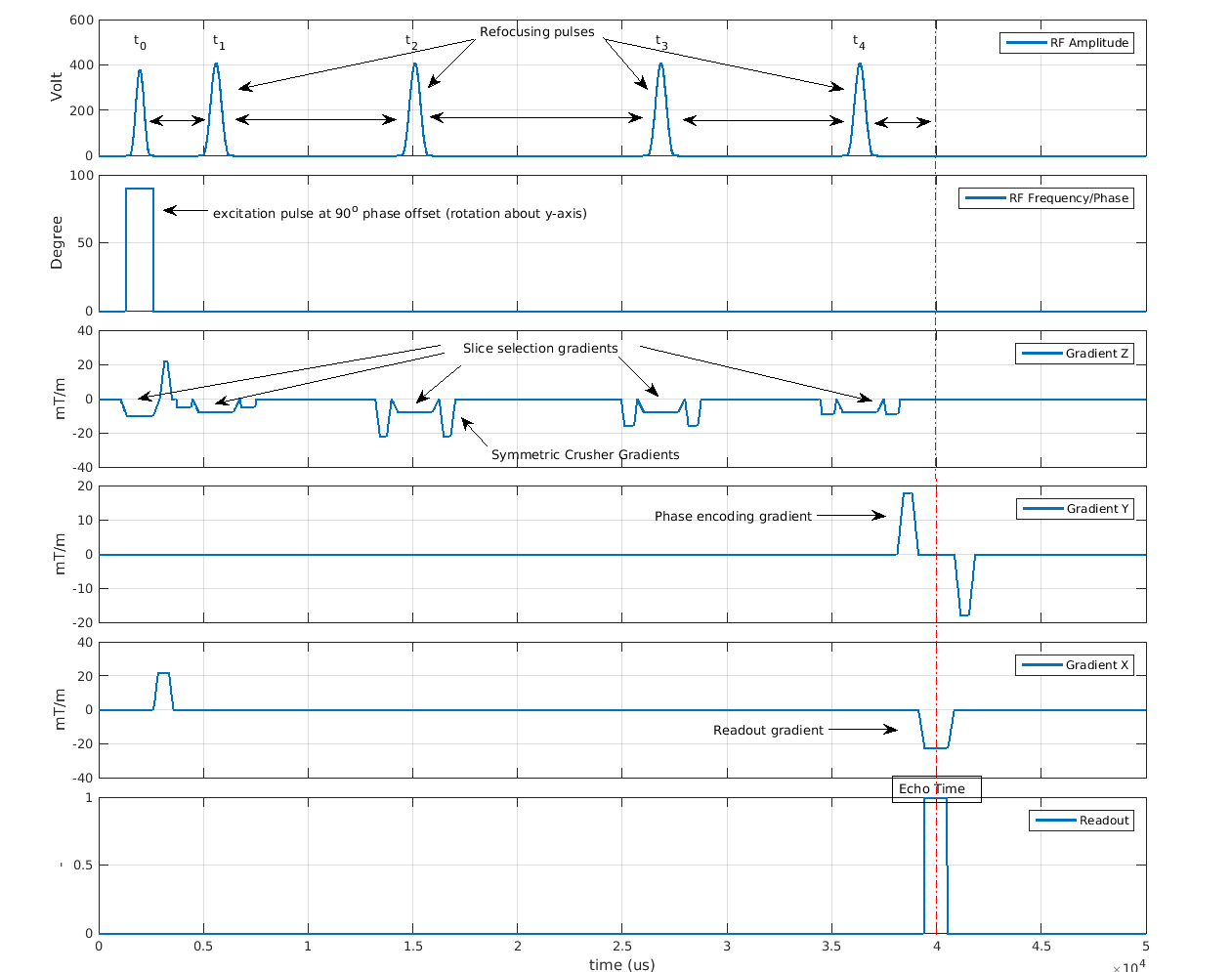}
\caption{Pulse sequence diagram for four-pulse \UDD\ sequence, which
  differs from the conventional \CPMG\ sequence in the spacing of the
  refocusing pulses.} \label{fig:seq_diagram}
\end{figure*}

The pulse sequence design consists of a slice-selective excitation pulse with a flip angle of $90^\circ$ about the $y$-axis to transfer magnetization into the transverse plane, followed by a series of \new{slice-selective} $180^\circ$ refocusing pulses about the $x$-axis, timed to coincide with \CPMG\ or \UDD\ spacing as defined in Eq.~(\ref{eq:UDD-timing}), followed by readout at the echo time $T_E$.  In addition to frequency and phase encoding gradients as required for imaging, the refocusing pulses are flanked by crusher gradients with variable gradient moments as shown in the sequence diagram Fig.~\ref{fig:seq_diagram}.  As the objective is to elucidate the effect of changes in the pulses timing, care was taken to ensure that the \CPMG\ and \UDD\ sequence variants were identical except for the differences in pulse timing, which was achieved by having a single implementation with a simple toggle switch to alternate between equal (\CPMG) and \UDD\ spacing of the refocusing pulses.

While the refocusing pulse sequence fixes the spacing of the refocusing pulses, it does not prescribe the excitation and refocusing pulses.  Ideal refocusing pulses should produce instantaneous $180^\circ$ rotations but this is clearly not possible as realistic pulses have finite width and there are constraints on the pulse amplitudes.  In addition the pulses are generally required to be slice-selective, which restricts the pulse shapes.  We choose Hanning-filtered \SINC\ pulses with pulse envelopes
\begin{equation}
   A(t) = \tfrac{1}{2} (1+\cos(2\pi t/t_{\RF})) \sinc(\pi c t/t_{\RF}), 
\end{equation}
for $ -\tfrac{1}{2} t_{\RF} \le t \le \tfrac{1}{2} t_{\RF}$ and  $c=2.7$ (fixed) where $t_{\RF}$ is the pulse duration, which can be dynamically adjusted in the sequence protocol.  As pulse sequences such as \UDD\ and \CPMG\ are designed for instantaneous pulses, shorter RF pulses, allowing shorter $T_E$, or more refocusing pulses for a given $T_E$, are desirable.  However, in practice the pulse durations are constrained by technical limitations of the equipment (e.g., available transmitter voltages) and deterioration in the quality of the slice profiles for very short pulses. 

\new{Following previous work in this area (e.g.~\cite{Warren2009}) only the final echo is acquired for both \CPMG\ and \UDD.  This departs from common practice in MRI, where multi-echo sequences such as \CPMG\ are mostly used to accelerate image acquisition in FSE sequences and for $T_2$ mapping, and usually all intermediate echoes are used.  Although it is possible to acquire intermediate echoes for the \UDD\ sequence, the non-periodicity of the refocusing pulse timings means that \UDD$_4$ is not equal to repeating \UDD$_2$ twice, and therefore the interpretation of the intermediate echoes is not obvious.  The inability to utilize intermediate echoes for \UDD\ $T_2$ mapping is a disadvantage in terms acquisition times.  However, this drawback can be offset by using slice selective refocusing pulses to perform $T_2$ mapping for multiple slices at once.}
  
\subsection{Phantom placement, coil and slice selection}

\begin{figure*}
 \subfloat[Coronal]{\includegraphics[width=0.35\textwidth]{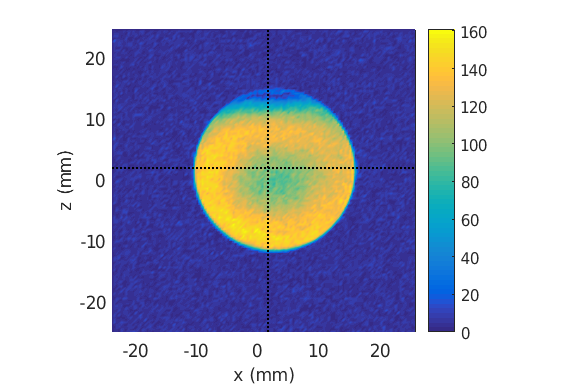}\hspace{-3ex}}
 \subfloat[Transverse]{\includegraphics[width=0.35\textwidth]{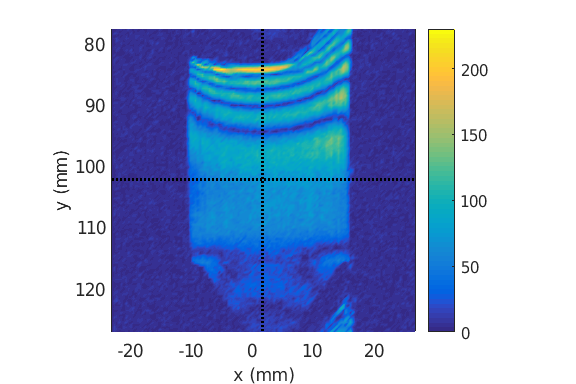}\hspace{-3ex}}
 \subfloat[Sagittal]{\includegraphics[width=0.35\textwidth]{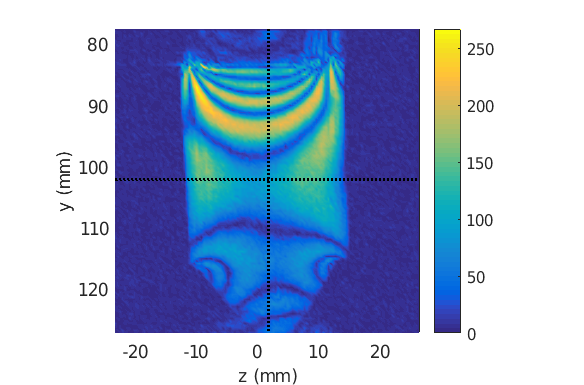}\hspace{-3ex}}
\caption{\CPMG\ field maps for 1\% agarose gel near ISO center. \label{fig:fields}}
\end{figure*}

The test-tube-sized agarose gels were placed in a cardboard sample holder on the patient table and positioned near the isocenter of the scanner.   The build-in transmit body coil was used for the excitation and refocusing pulses, while a \SI{7}{cm} loop coil was used as the receive coil.  For all gels a \SI{5}{mm} coronal slice was selected as shown in Fig.~\ref{fig:fields}.   Field-mapping was used to select the slice to avoid the field inhomogeneities near the top and bottom of the sample.  The images acquired were automatically registered using in-house \texttt{matlab} code and circular regions of interest (ROI) selected.  The radius of the ROI was about 80\% of the phantom radius to avoid edge effects near the boundary.

\begin{figure*}
  \hspace{-6ex}
\subfloat[ROIs]{\includegraphics[width=0.35\textwidth]{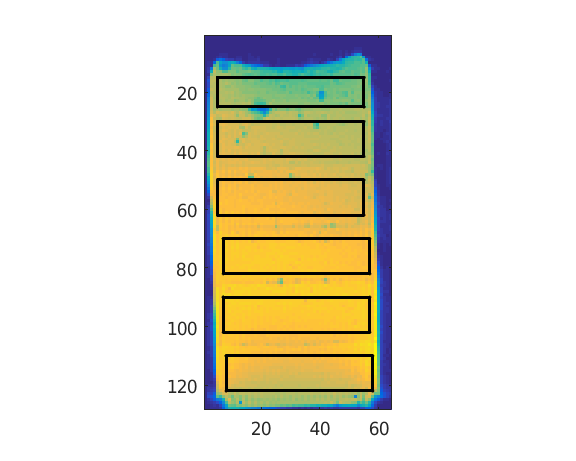}\hspace{-8ex}}
\subfloat[Coronal]{\includegraphics[width=0.35\textwidth]{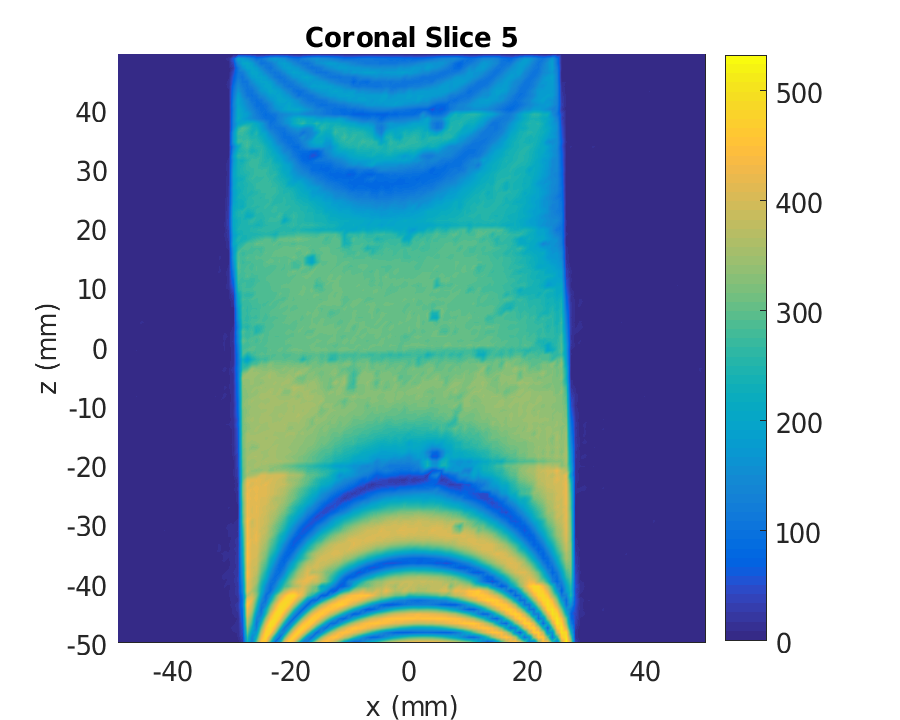}\hspace{-2ex}}
\subfloat[Sagittal]{\includegraphics[width=0.35\textwidth]{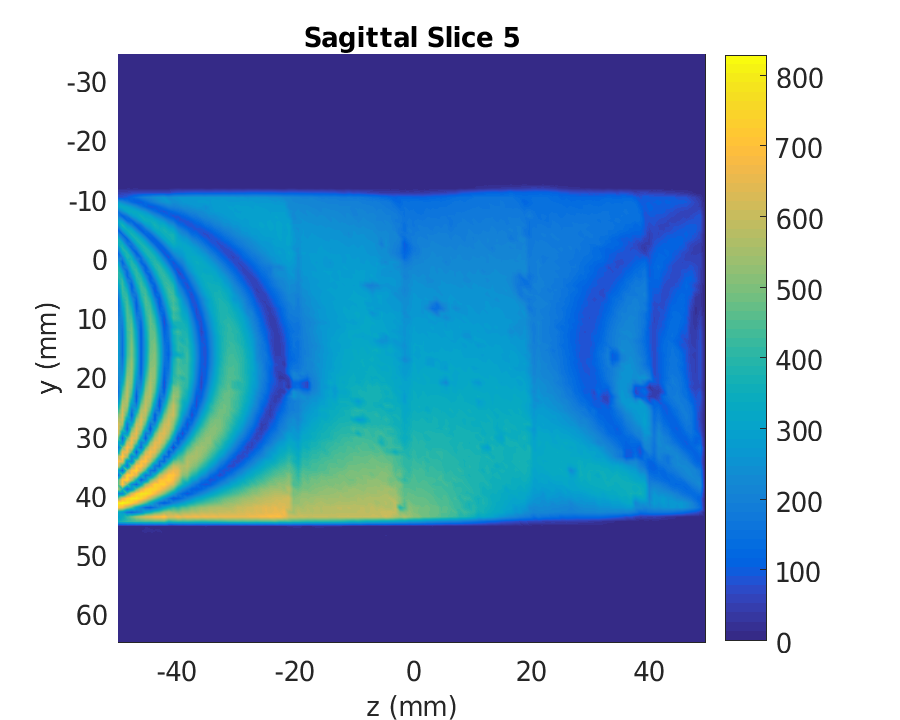}}
\caption{In-slice ROI selection and \CPMG\ field maps for six-layer agar gel phantom. \label{fig:fields-agar}}
\end{figure*}

The larger cylindrical agar hydrogel phantom was placed in a cylindrical knee coil aligned with the $z$-axis approximately \SI{15}{mm} below the isocenter of the magnet.  Unlike the loop coil used for the small samples, the knee coil is a transmit-receive birdcage resonator and was thus used both for RF excitation and readout.  A \SI{5}{mm} coronal slice through the center of the phantom was selected and six rectangular ROI were chosen as shown in Fig.~\ref{fig:fields-agar}.

\subsection{Transmitter Voltage Calibration}

\begin{figure}
  \center \includegraphics[width=\columnwidth]{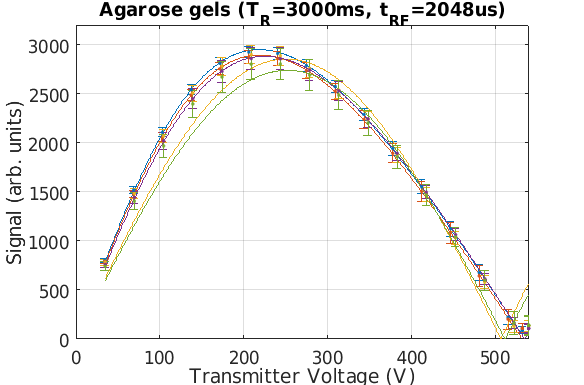}
  \caption{Signal vs transmitter voltage for \SI{2048}{\micro\second} SINC pulses for test-tube-size agarose gels using body coil for transmit and loop coil for detection.}
\label{fig:flip-angle1}
\end{figure}

To determine the transmitter voltages required to obtain the desired flip angles, a number of gradient echo scans with $T_E=\SI{3}{\milli\second}$ and $T_R=\SI{3000}{\milli\second}$, respectively, and varying transmitter voltage $V$ were performed.  The mean signal $S(V)$ over a selected ROI of each sample was measured for each scan, and the signal $S(V)$ fitted according to the formula
\begin{equation}
\label{eq:Signal}
  S(V) = c_0  \frac{|\sin (c_1 V)|}{ 1-e^{-T_R/T_1} \cos (c_1 V)}.
\end{equation}
The resulting signal fits yield the flip angle as a function of the voltage, $\alpha(V) = c_1 V$, assuming linear dependence, as well as the constants $c_0$ related to the saturation magnetization and the longitudinal relaxation time $T_1$ as a bonus.  The results of the flip angle characterization for our agarose gels subject to \SINC\ pulses of length $t_{\RF}=\SI{2048}{\micro\second}$ are shown in Fig.~\ref{fig:flip-angle1}.   We infer from the graph that the voltage required to obtain a $180^\circ$-pulse of length $t_{\RF}=\SI{2048}{\micro\second}$ is around $\SI{535}{V}$.  Notice the signal curves for all agarose gels are virtually coincident with small errorbars indicating little variation of the signal over the ROI.

\begin{figure}
  \includegraphics[width=\columnwidth]{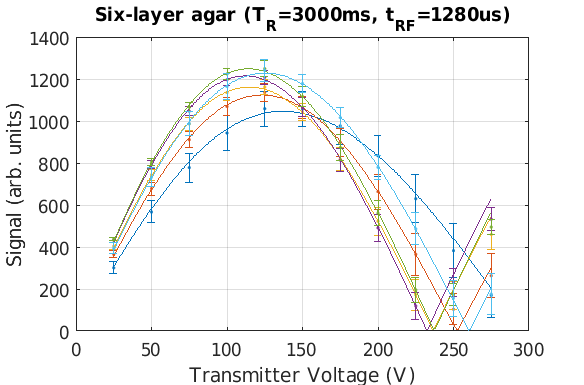}
  \includegraphics[width=\columnwidth]{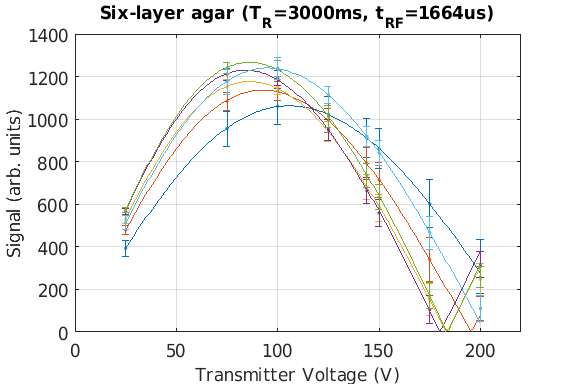}
  \caption{Signal vs transmitter voltage for \SI{1280}{\micro\second} and \SI{1664}{\micro\second} SINC pulses for six-layer agar gel using 15 channel knee coil as transmit and receive coil.}
\label{fig:flip-angle2}
\end{figure}

The equivalent signal vs transmitter voltage curves for the six ROIs of the six-layer agar gel phantom defined above, are shown in Fig.~\ref{fig:flip-angle2}.   Due to RF in homogeneity there is much greater variation of the signal over each ROI and the flip angle curves and the voltages required to achieve a $180^\circ$-pulse.  This means that it is impossible to choose a transmitter voltage that achieves uniform flip angles over the entire phantom.

\begin{figure*}
\center\includegraphics[width=0.7\textwidth]{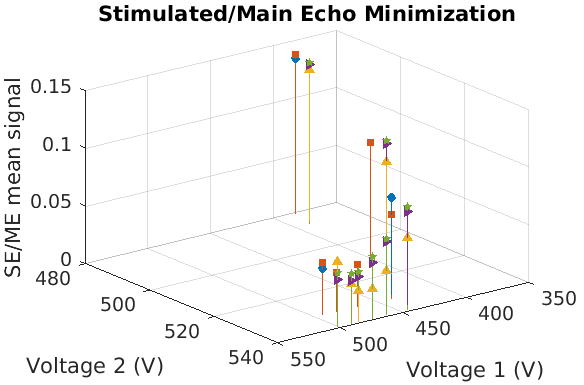}
\caption{Stimulated echo vs main echo signal for two-pulse \CPMG\ sequence as a function of the voltages of the \RF\ excitation ($V_1$) and refocusing pulses ($V_2$) for the test-tube sized agarose gels scanned in the loop coil.  The voltages are tuned to minimize the stimulated echo contributions.  For our choice of \RF-pulse of durations \SI{1280}{\micro\second} and \SI{2048}{\micro\second}, respectively, we attained SE/ME ratios between 2.7\% and 4.4\%.  (Different markers correspond to different gel samples.)}
\label{fig:SE_vs_ME}
\end{figure*}

To further calibrate the pulses to minimize unwanted stimulated echoes, a number of \CPMG\ experiments were performed.  Each experiment was carried out with two choices for the crusher gradients flanking the refocusing pulses: (1) symmetric crushers with gradients moment $g_{1a}=g_{1b}$ and $g_{2a}=g_{2b}\neq g_{1a}$, which suppress the stimulated echo, and (2) $g_{1a}=g_{2b}$ and $g_{1b}=g_{2a}=0$, which dephase the regular echo and select the stimulated echo, where $a$ and $b$ refer to the left and right-flanking crushers, respectively. The voltages were then tuned to minimize the ratio of the stimulated echo and primary echo signals.  Fig.~\ref{fig:SE_vs_ME} shows the ratios of the signal intensity for the stimulated and regular echo, averaged over the ROI, for the test-tube sized agarose gels as stem plots versus the peak voltage of the excitation and refocusing pulse.  The data suggests that the optimal voltages for all samples cluster around \SI{500}{V} and \SI{535}{V} for the excitation and refocusing pulses, respectively.

\begin{figure*}
  \includegraphics[width=0.26\textwidth]{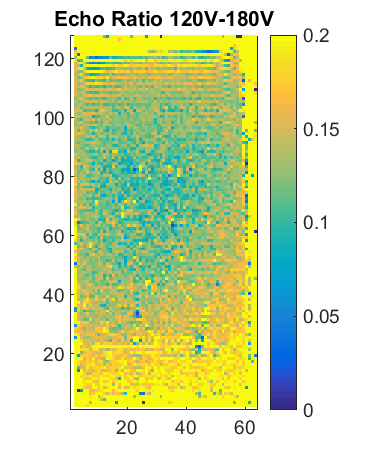}\hspace{-4ex}
  \includegraphics[width=0.26\textwidth]{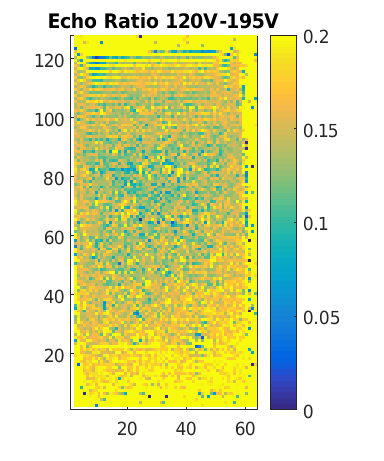}\hspace{-4ex}
  \includegraphics[width=0.26\textwidth]{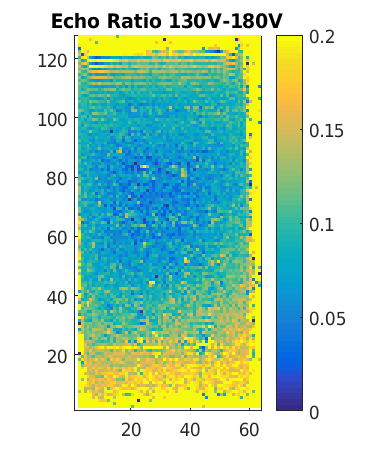}\hspace{-4ex}
  \includegraphics[width=0.26\textwidth]{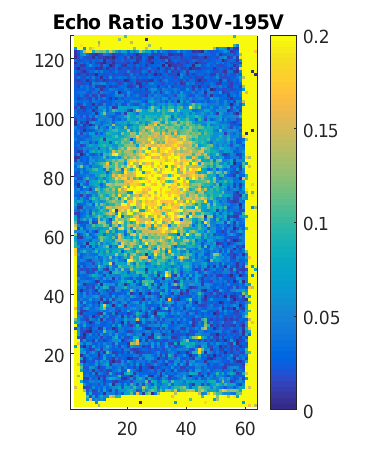}
  \caption{Stimulated echo vs main echo signal for two-pulse \CPMG\ sequence for four different combinations of peak pulse amplitudes for the six-layer agar gel phantom.}
\label{fig:SE_vs_ME2}
\end{figure*}

For the six-layer agar gel phantom RF inhomogeneity leads to considerable variation in the stimulated vs regular echo ratios across the phantom as shown in Fig.~\ref{fig:SE_vs_ME2} and the optimal choice of voltages for the excitation and refocusing pulses is less obvious.  We choose \SI{130}{V} and \SI{180}{V} for the amplitudes of the excitation and refocusing pulses, respectively, because it resulted the most uniform SE/PE ratio compared to other choices that resulted in lower stimulated echo contributions for some parts of the phantom but higher values elsewhere.

\subsection{Transverse relaxation subject to \CPMG$_n$ and \UDD$_n$}

\begin{figure*}
  \includegraphics[width=0.33\textwidth]{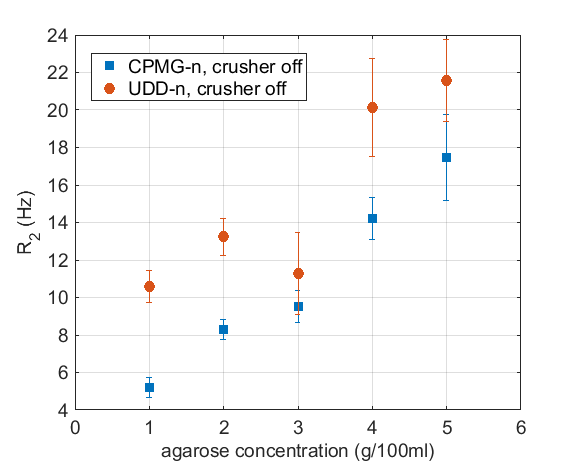}\hspace{-3ex}
  \includegraphics[width=0.33\textwidth]{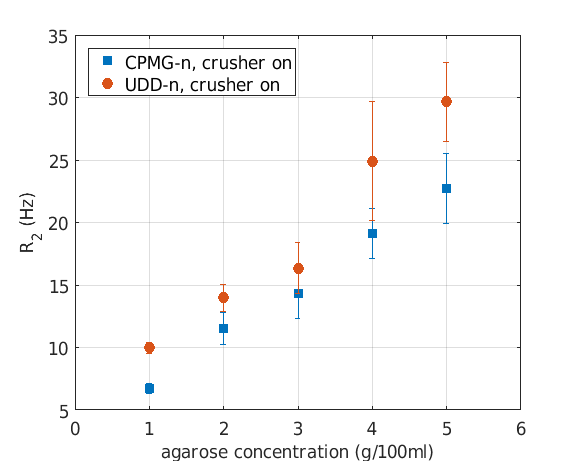}\hspace{-3ex}
  \includegraphics[width=0.33\textwidth]{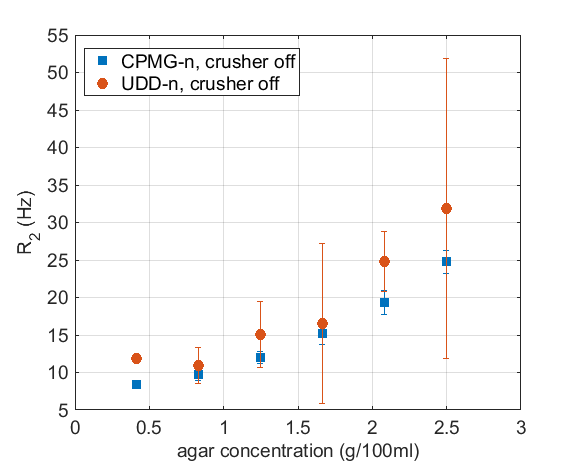}
  \caption{$R_2$ for agarose samples and six-layer agar gel phantom obtained used \CPMG$_{n}$ and \UDD$_{n}$ by fitting the signal according to \eqref{eq:S_TE} as shown in Figs~ \ref{fig:T2-fits1}, \ref{fig:T2-fits2} and \ref{fig:T2-fits3} in the appendix.}
  \label{fig:R2-summary}
\end{figure*}

To determine the transverse relaxation rates for our phantoms subject to different decoupling sequences, a series of images were acquired for $n$-pulse \CPMG\ and \UDD\ sequences with $n$ even, ranging from $2$ to $16$.  The mean signal $S(T_E)$ over a fixed ROI was measured for each sequence and fitted according to
\begin{equation}
  \label{eq:S_TE}
  \ln S(T_E) = C - T_E R_2
\end{equation}
in both cases to obtain $R_2^{\CPMG}$ and $R_2^{\UDD}$.  The echo times $T_E(n)$ were the same for both \CPMG$_{n}$ and \UDD$_{n}$.  For the agarose gels we compared two cases: (a) no crushers, \textit{i.e.}, crusher gradient moments zero and (b) symmetric crusher gradients with unequal moments designed to suppress all but the target pathway echo.  For the six-layer agar gel phantom only the results without crusher gradients were used for $R_2$ fitting due to low signal values with crusher gradients applied, as expected considering the large stimulated echo contributions for this phantom.  The results are shown in \ref{app:T2} and summarized in Fig.~\ref{fig:R2-summary}.  As expected, we observe an almost linear increase in $R_2$ with concentration in all cases.  Comparison of the $R_2$ values for \CPMG\ and \UDD, however, shows that \CPMG\ consistently outperforms \UDD\ in the sense that the observed relaxation rates with \CPMG\ decoupling are consistently lower than for \UDD\ decoupling.  \new{The $R_2$ fits for \CPMG\ and \UDD\ with uniform crusher gradients in Fig.~\ref{fig:T2-fits4} in the Appendix, further show that the difference between \CPMG\ and \UDD\ is negligible for the middle layers (3 and 4), for which the field maps show high $B_0$ homogeneity, while the \CPMG\ performs significantly better than \UDD\ for the outer layers, which are subject to significant $B_0$ inhomogeneity.}

\new{Although the transverse relaxation rates $R_2=1/T_2$ obtained from our fits are likely to be overestimated due to fractional signal losses for each refocusing pulse, the comparison should still be valid as these losses should be the same for both \UDD\ and \CPMG, considering that they differ only in the timing of the refocusing pulses.  However, there are a number of ways the sequence could be improved.  $T_2$ underestimation due to fractional signal loss could be corrected using $B_1$ mapping \cite{Poon1992} and by optimizing the refocusing angles to maximize in final echo amplitudes~ \cite{Henning2001}.}
  
\section{Dynamic Decoupling Simulations for Ideal Pulses}

\begin{figure*}
  \subfloat[Pink noise: $1/f$ spectrum]{\includegraphics[width=0.32\textwidth]{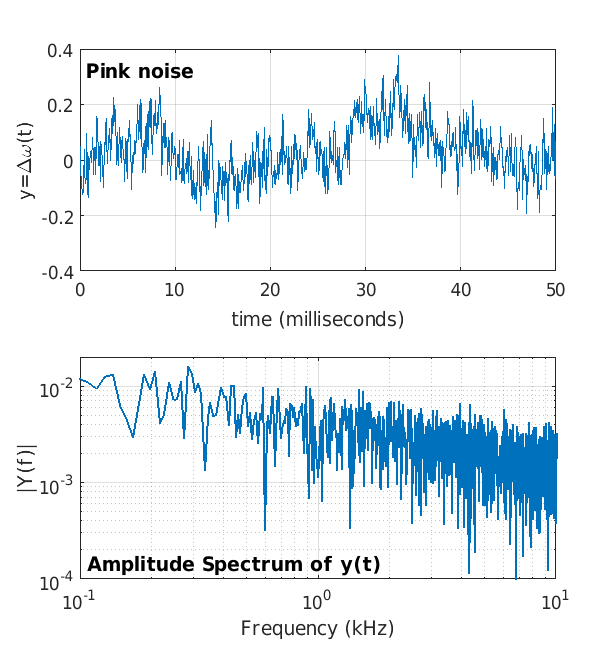}}
  \subfloat[Noise noise: flat spectrum]{\includegraphics[width=0.32\textwidth]{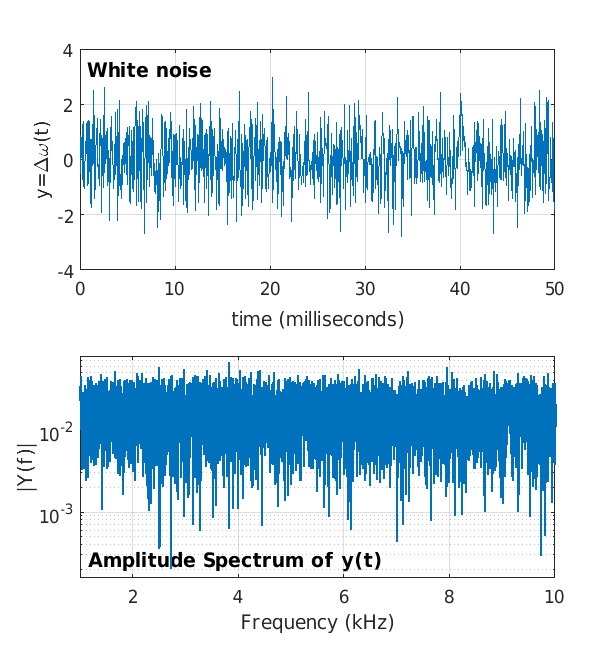}} 
  \subfloat[Blue noise: $~f$ spectrum]{\includegraphics[width=0.32\textwidth]{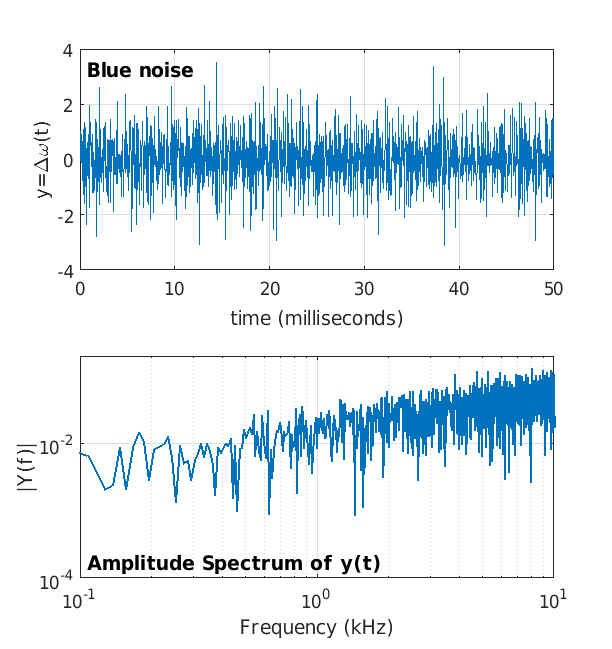}}
  \caption{Examples of resonance frequency fluctuations $\Delta\omega(t)$ for different noise types.}\label{fig:noise1}
\end{figure*}

To explain these results, we performed simulations comparing the performance of ideal dynamic decoupling sequences with \CPMG\ and \UDD\ spacing, respectively, for different noise spectra.  Assuming (i) that we start with all the magnetization transferred to the $xy$ plane, $M_z(0)=0$, and (ii) that transverse magnetization loss due to $T_1$ relaxation is negligible on the relevant time scale, \textit{i.e.}, $T_1\gg T_E$, we can assume $M_z(t) \approx 0$ for $0\le t\le T_E$ and reduce the Bloch equation to a single equation for the transverse magnetization $M_{xy}=M_x+iM_y$,
\begin{equation}
  \dot{M}_{xy}(t) = -\gamma(t) M_{xy}(t)
\end{equation}
with $\gamma(t)= i\omega(t)+T_2^{-1}$.  Integrating this equation yields the explicit formula
\begin{equation}
   M_{xy}(t) = \exp \left[-\int_0^t \gamma(t) dt \right] M_{xy}(0)
\end{equation}
for the transverse magnetization at time $t$.  To investigate the effect of different noise spectra, we set $\Delta\omega(t) = c_0+\eta(t)$, where $\eta(t)$ is either noise with a flat frequency spectrum (white noise), noise with a spectral density proportional to $1/f$ (pink noise) or noise a spectral density proportional to $f$ (blue noise).  For each case, 10,000 instances  of $\eta(t)$ are generated.  As we are not interested in the constant offset here, which is refocused by a single spin echo, we set $c_0=0$.  We also drop the $1/T_2$ term, which is independent of the dephasing resulting from resonance frequency fluctuations.  In our simulations there was no explicit high-frequency cut-off but the frequency resolution is limited by the time resolution of the noise sampling of \SI{50}{\micro\second} and the length of the signal of \SI{50}{\milli\second}.  Examples of the different types of noise in the the time and frequency domain are shown in Fig.~\ref{fig:noise1}.  Note that in a slight departure from convention we are applying the terms pink, white and blue to amplitude spectra, not the power spectral density, as the former are directly related to the induced dephasing.  The evolution of the magnetization $M_{xy}^\eta(t)$ is calculated for each $\eta(t)$ and the average magnetization computed by averaging the trajectories.

\begin{figure*}
  \subfloat[Pink noise: $1/f$ spectrum]{\includegraphics[width=0.32\textwidth]{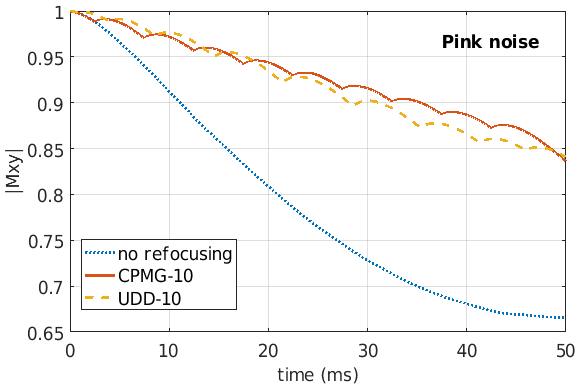}}
  \subfloat[White noise: flat spectrum]{\includegraphics[width=0.32\textwidth]{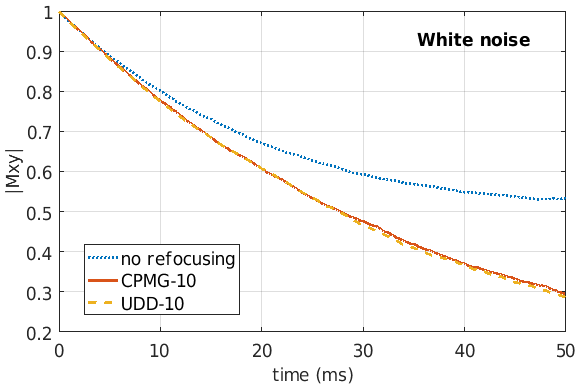}}
  \subfloat[Blue noise: $f$ spectrum]{\includegraphics[width=0.32\textwidth]{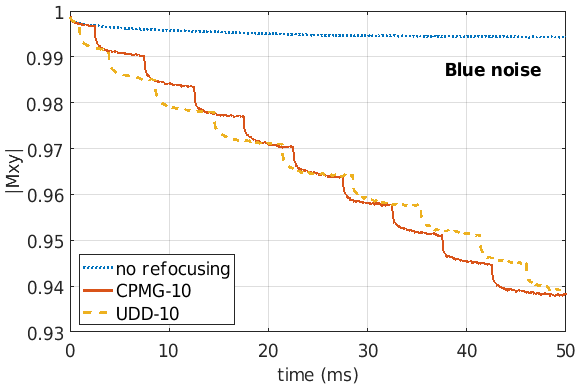}}
  \caption{Evolution of transverse magnetization subject to different noise models averaged over $10^4$ noise
  trajectories without refocusing and ten pulse \CPMG\ and \UDD\ refocusing sequences.}\label{fig:noise2}
\end{figure*}

\begin{figure*}
  \subfloat[Pink noise: $1/f$ spectrum]{\includegraphics[width=0.32\textwidth]{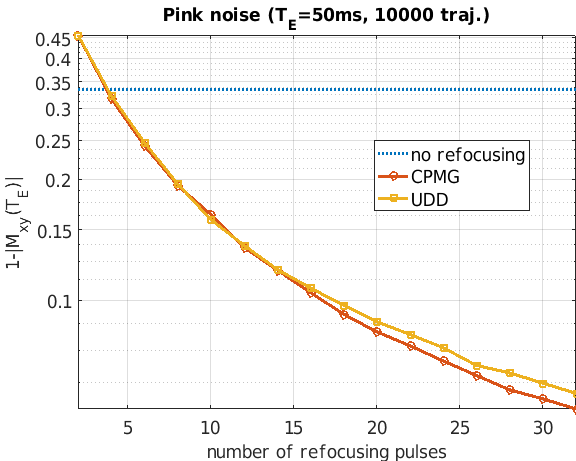}}
  \subfloat[White noise: flat spectrum]{\includegraphics[width=0.32\textwidth]{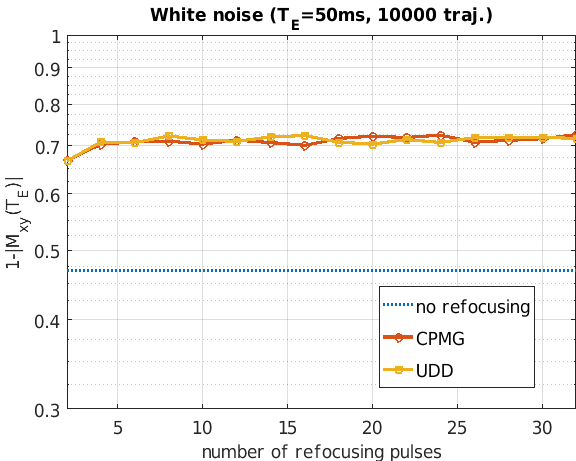}}
  \subfloat[Blue noise: $f$ spectrum]{\includegraphics[width=0.32\textwidth]{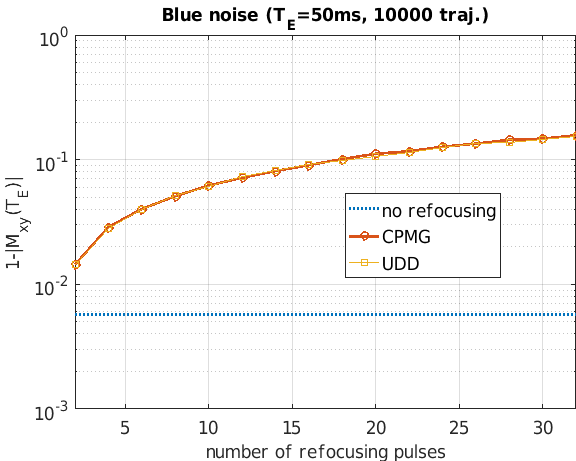}}
  \caption{Magnetization loss $1-|M_{xy}(t)|$ at the echo time $T_E=\SI{50}{ms}$ as a function of
    the number of refocusing pulses $n$ for different noise models.}\label{fig:noise3}
\end{figure*}

Fig.~\ref{fig:noise2} shows that the refocusing pulses are effective for pink noise in that the average transverse magnetization $|M_{xy}(t)|$ reduces more slowly subject to refocusing pulses.  Fig.~\ref{fig:noise3} further shows that the effect increases with the number of refocusing pulses, as expected.  However, there is little difference in $|M_{xy}|$ between \UDD\ and \CPMG.  In fact the transverse magnetization loss for \CPMG\ is slightly smaller than for \UDD.  For blue noise \UDD\ appears to have a slight advantage over \CPMG\ for $n=10$ but comparing the magnetization loss subject to either \CPMG\ or \UDD\ refocusing and without refocusing for white and blue noise shows that refocusing is actually detrimental in this case in that it accelerates the loss of transverse magnetization.  For white noise the effect appears independent of the number of refocusing pulses, while for blue noise the magnetization loss increases with the number of refocusing pulses.

These results are not too surprising considering that the cumulative effects of high frequency fluctuations on the phases of the proton spins tend to cancel over longer periods of time, and the longer the time averaged over, the better the averaging of the high-frequency components.  On the other hand, if noise is dominated by low frequencies as in the case of pink ($1/f$) noise, then the echo time $T_E$ is too short for averaging to occur and refocusing becomes beneficial.  They are also consistent with other theoretical work.

For example, \cite{Cywinski2008} found that for $1/f$ and random telegraph noise, \UDD\ is optimal for suppressing initial decoherence in the presence of a hard ultraviolet cut-off in the Gaussian noise spectrum, but if the cut-off cannot be reached then \CPMG\ is the better practical approach.  The results were corroborated in \cite{Lange2010, Biercuk2009} and Pasini and Uhrig noted that for a pure-dephasing spin boson model for baths with power law spectra, the numerical optimization yielded dynamic decoupling sequences very close to \CPMG\ for bath spectra with a soft cut-off but the solutions approached \UDD\ as the cut-off became harder \cite{Pasini2010}.

Ref.~\cite{Biercuk2009} examined the efficacy of dynamic decoupling sequences with regard to phase error suppression applied to a model quantum memory.  The ambient magnetic field fluctuations in the high-field superconducting magnet were measured directly, showing a $1/f^{2}$ spectrum (power spectrum $1/f^{4}$) with a soft high-frequency cut-off.  \CPMG\ was found to perform similarly to \UDD.  Numerical simulations with artificially injected noise showed that \UDD\ significantly outperformed \CPMG\ when Ohmic noise with a high-frequency cut-off was injected, whereas \CPMG\ performed similarly to \UDD\ when $1/f$ noise with a high-frequency cut-off was injected.

\section{Experimental noise spectra for gel phantoms and tissue}

\begin{figure*}
  \subfloat[Noise spectrum: gel]{\includegraphics[width=0.35\textwidth]{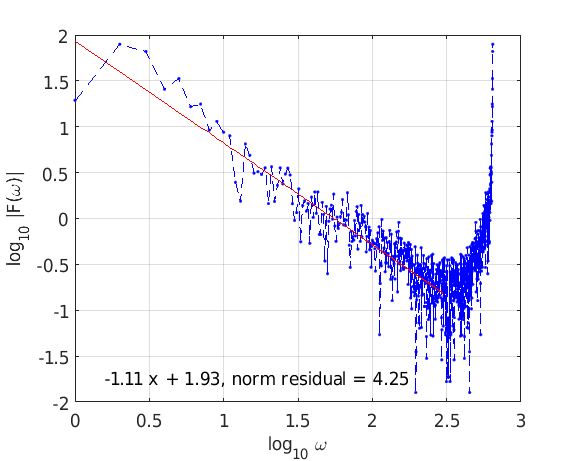}
    \hspace{-3ex}}
  \subfloat[Noise spectrum: prostate]{\includegraphics[width=0.35\textwidth]{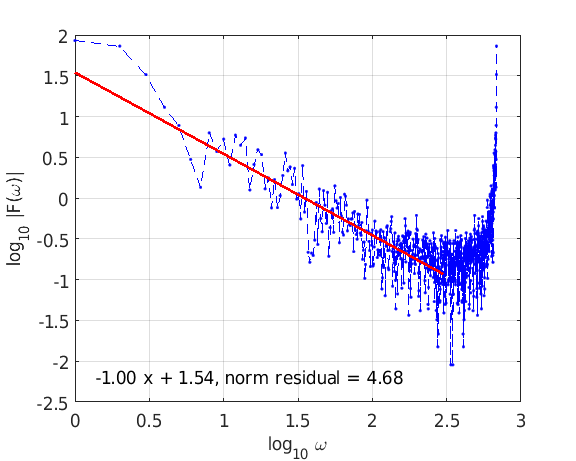}
    \hspace{-3ex}}
  \subfloat[Histogram of $\alpha$]{\includegraphics[width=0.35\textwidth]{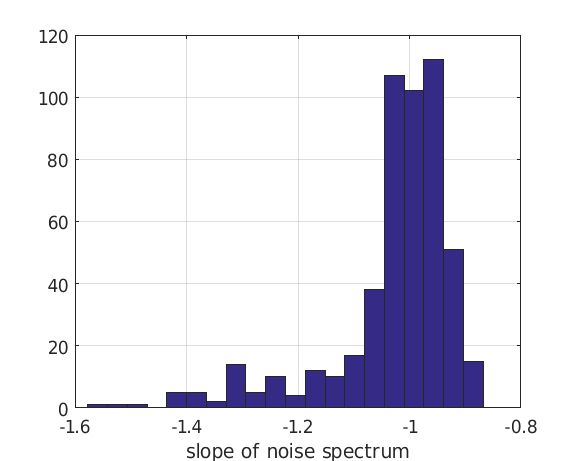}}
  \caption{Examples of experimental noise spectra for a gel phantom and prostate show that the amplitude noise spectrum has an $f^\alpha$ dependence with $\alpha$ around $-1$ and that the experimental noise spectra for our gel phantoms closely resemble those of human tissue.  The histogram plot of $\alpha$, i.e., the slopes of the noise spectra in the log-log plot, derived from 512 prostate spectra shows a distribution clustered about $-1$, as expected for pink noise.}\label{fig:noise-exp}
\end{figure*}

To assess which noise model best describes the noise spectra for tissue and tissue-mimicking gel phantoms, we obtained experimental noise spectra for both using standard spectroscopy sequences for point-resolved single voxel spectroscopy and chemical shift imaging (CSI), \textsc{PRESS} and \textsc{sLASER}. The noise spectra were obtained from the normal spectra following the technique described in \cite{Relano2002}. For the prostate data informed consent was obtained from a volunteer under the centre's ethical approval for pilot studies.

Fig.~\ref{fig:noise-exp} shows that the amplitude noise spectra derived from experimental spectra for a gel phantom and prostate \textit{in vivo} scans strongly resemble pink noise spectra.  More specifically, they resemble pink noise spectra with a high frequency spike near the frequency cut-off determined by the sampling rate of the receive coils.  Thus, the noise spectra are almost opposite of blue noise with a high-frequency cut-off.  Therefore, our empirical observation that \CPMG\ provides better refocusing in tissue-mimicking gels makes sense, and given the similarity between the noise spectra of our gels and \textit{in vivo} spectra for human tissue, it stands to reason that the same results are likely to apply for tissue imaging.  Pink noise also appears to be a more realistic noise model for biological systems and tissue-mimicking gels on theoretical grounds, considering that fluctuations due to $J$-coupling and diffusive motion effects are typically low-frequency, on the order of a few hundred Hertz at most.  

However, the simulations do not explain the empirical data fully.  The differences in the observed signal strength between \UDD\ and \CPMG\ appear larger than the small differences observed in the simulations.  Furthermore, the simulations do not explain the apparent enhancement of tissue inhomogeneities observed in the literature~\cite{Warren2009}.  To elucidate these observations requires more in depth analysis considering the effects of various non-idealities that can result in imperfect excitation and refocusing pulses.

\section{Non-idealities and Coherence Pathways Analysis}

An ideal excitation pulse converts the longitudinal magnetization of the entire slice uniformly to transverse magnetization and a perfect refocusing pulse inverts the sign of this transverse magnetization.  Realistic excitation and refocusing pulses, however, are not perfect for various reasons ranging from inevitable variations across slices due to non-rectangular pulse profiles, especially for short excitation and refocusing pulses, to $B_0$ and \RF\ inhomogeneity either due to technical imperfections in the equipment or tissue susceptibility effects, for example.

\begin{figure*}
\includegraphics[width=0.5\textwidth]{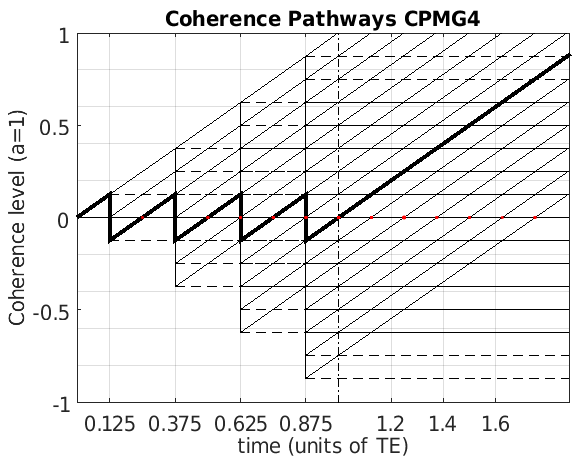}\hfill
\includegraphics[width=0.5\textwidth]{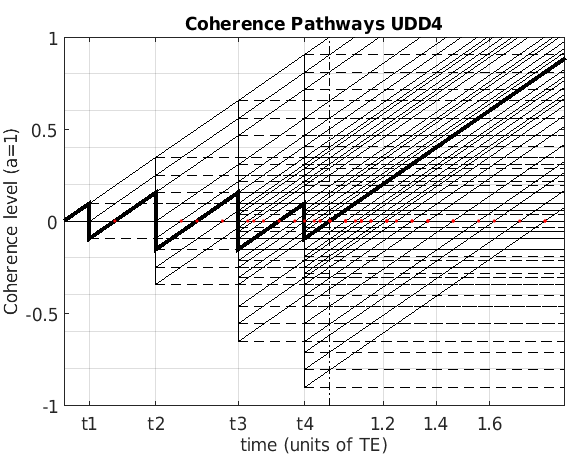}
\caption{Possible coherence pathways for an excitation pulse followed by a four refocusing pulses with \CPMG$_4$ spacing (left) and \UDD$_4$ spacing (right) with dephasing rate $a=1$ per echo period $T_E$.  $t_n$ are the \UDD$_4$ timings determined by (\ref{eq:UDD-timing}). The thick line indicates the target pathway.  Red dots indicate possible echoes.}
\label{fig:pathways}
\end{figure*}

To elucidate the effects of applying a sequence of imperfect pulses on the evolution of an ensemble of spins we use coherence pathways analysis~\cite{Scheffler1999}.  Let $M_z$ be the longitudinal magnetization and decompose the transverse magnetization $M_t$ into dephasing and rephasing components, $M_+=M_x+iM_y$ and $M_-=M_x-iM_y$.  An instantaneous pulse effecting a rotation $R$ about a general axis maps the magnetization $\vec{M} = [M_x,M_y,M_z]^T$ to the new magnetization vector $\vec{M}^+$ by a linear transformation $\vec{M}^+ = R \vec{M}$ with
\begin{equation}
   \label{eq:R}
    R = \exp \left[ \alpha \begin{pmatrix} 
                        0 &         -\cos\theta &  \sin\phi\sin\theta\\
               \cos\theta &                   0 & -\cos\phi\sin\theta\\
      -\sin\phi\sin\theta & \cos\phi\sin\theta  &                 0
    \end{pmatrix} \right]
\end{equation}
for a rotation about the axis $(\sin\theta\sin\phi,\sin\theta\cos\phi,\cos\theta)$, where $\theta$ and $\phi$ are the polar and azimuthal angles, respectively. Substituting $M_x = \tfrac{1}{2}(M_++iM_-)$ and $M_y = -\tfrac{i}{2}(M_+-iM_-)$ we can easily derive the corresponding linear transformation $A$ mapping the pre-pulse coherence vector $\vec{F} = [M_+,M_-,M_z]$ to the post-pulse vector $\vec{F}^+$.  Each pulse splits the ensemble into three components, which evolve between pulses according to
\begin{equation} 
 \label{eq:evolve}
  M_+ = e^{-\tau/T_2} M_+, \quad
  M_- = e^{-\tau/T_2} M_-, \quad
  M_z = e^{-\tau/T_1} M_z,
\end{equation}
where $\tau$ is the time elapsed since the last pulse and $T_1$ and $T_2$ are the longitudinal and transverse relaxation rates, respectively.  Combining these results enables us to compute the amplitudes of the coherences $M_+$, $M_-$ and $M_z$ for all coherence pathways for a multi-pulse sequence of instantaneous rotations and delays.  The coherence pathways and echo formation can be visualized diagrammatically as shown in Fig.~\ref{fig:pathways}.

We can label the pathways by the action of the pulses, using $0$ if the pulse does not change the magnetization, $1$ if it results in a $90^\circ$ rotation and $2$ if it results in a $180^\circ$ rotation.  With this labeling of the pathways, careful analysis of the pathway diagrams shows that for constant dephasing the pathways 12222, 12020, 10202, 11001, 11221, 10110, 11111, 11122, 12112, 11012, 12101, 12211 are simultaneously refocused at the echo time $T_E$ for \CPMG$_4$, while for \UDD$_4$ pulse timing, four distinct echoes are formed around $T_E$: pathway 11122 forms an echo at time $0.9410T_E$, pathways X1211, XX120 with $X\in \{0,2\}$ form an echo at time $0.9635T_E$, pathways 12222, 11001, 11221, 10110, 12112 form an echo at time $T_E$, and pathways 11012, 12101, 12211 form an echo at time $1.0590T_E$.

\begin{figure*}
\subfloat[Ideal pulses, no relaxation]{\includegraphics[width=0.34\textwidth]{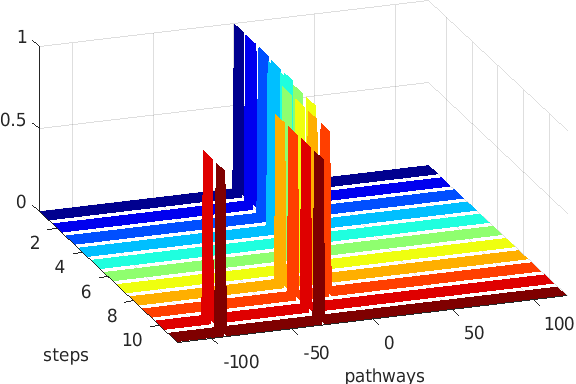}\hspace{-3ex}}
\subfloat[Ideal pulses, relaxation]{\includegraphics[width=0.34\textwidth]{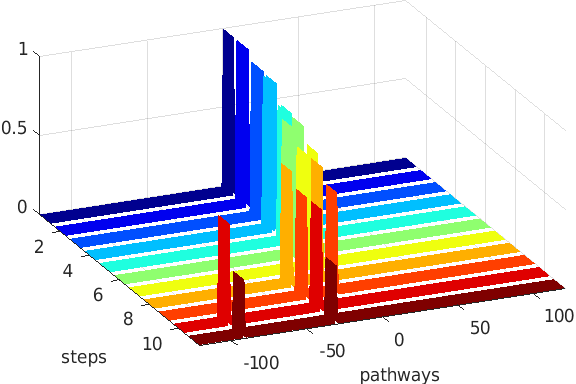}\hspace{-3ex}}
\subfloat[non-ideal pulses, relaxation]{\includegraphics[width=0.34\textwidth]{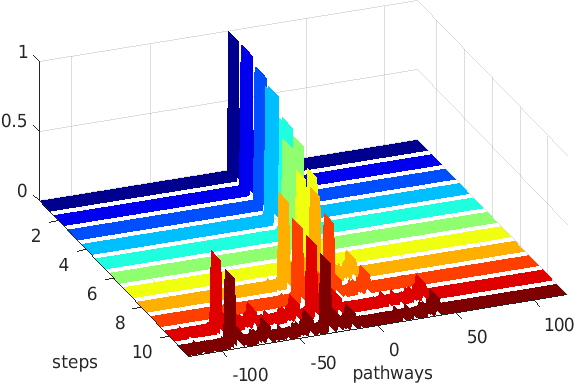}}
\caption{Evolution of coherence amplitudes for an excitation pulse followed by a four refocusing pulses, starting with the entire magnetization in $M_z$, for ideal pulses without relaxation (a), ideal pulses with relaxation (b) and non-ideal pulses with relaxation (c).  For typical $T_1$ and $T_2$ values and echo times the difference in the coherence amplitudes due to pulse spacing effects between \UDD$_4$ and \CPMG$_4$ is negligible.}
\label{fig:amplitudes}
\end{figure*}

If all pulses are perfect then the only coherence pathway excited by the four-pulse \CPMG\ or \UDD\ sequence is the pathways shown in bold in Fig.~\ref{fig:pathways}.  This is reflected in coherence amplitude plot in Fig.~\ref{fig:amplitudes} (a) showing that there are only two non-zero amplitudes at each step in the sequence, corresponding to the $M_+$ and $M_-$ magnetization of the target pathway.  With relaxation between pulses the amplitudes decay (b) but there are still only two matched peaks corresponding to $M_+$ and $M_-$ magnetization of the target pathway.  Any rotation axis or angle error, however, leads to population of many other coherence pathways as shown in Fig.~\ref{fig:amplitudes}(c).

Crusher gradients can be applied to selectively dephase coherence pathways by ensuring that their total gradient moments are non-zero and preferably large.  Our implementation allows crusher gradients with arbitrary gradient moments $a_{n,L}$ and $a_{n,R}$ before and after the $n$th refocusing pulse but to preserve the target pathway, the crusher gradient moments before and after each refocusing pulse must be equal $a_{n,L}=a_{n,R}=:a_n$.  Uniform crusher gradient moments $a_n=a$, \new{the default choice for the standard \CPMG\ sequence~\cite{Bernstein2004},} suppress unwanted FID signals but still refocus many stimulated echo pathways.  Dephasing the latter requires the crusher gradient moments for different refocusing pulses to be mutually distinct $a_n\neq a_{n'}$ for $n \neq n'$.  For $n=4$ the explicit conditions that should be satisfied are
\begin{align*}
  (a_2-a_1) \neq (a_4-a_3) \\
  (a_3-a_1) \neq (a_1-a_4) \\
  (a_4-a_1) \neq (a_2-a_4) \\
  2(a_1-a_4) + (a_2-a_3) & \neq 0.
\end{align*}
For sequences with more refocusing pulses the number and complexity of the equations increase dramatically.

\begin{figure}
\center\includegraphics[width=\columnwidth]{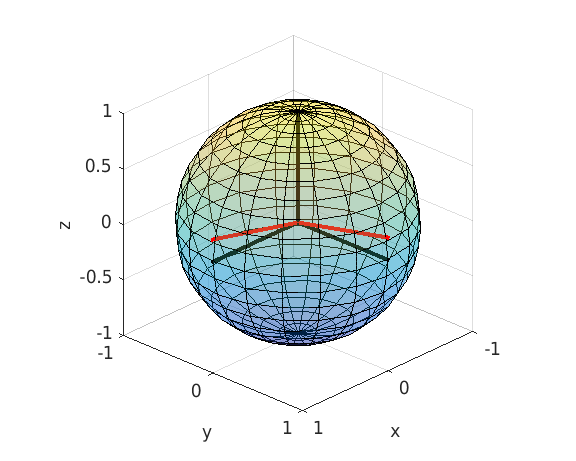}
\caption{Bloch sphere with $x$, $y$ and $z$-axis in black.  An off-resonant excitation pulse about the $x$ or $y$ axis will instead produce a rotation about one of the red axes.  The azimuthal angle of the actual rotation axis depends on the magnitude of the resonance offset $\Delta\omega$ relative to the Rabi frequency of the pulse, $\theta=\tfrac{\pi}{2}-\arctan(\Delta\omega/\Omega)$.  The generalized Rabi frequency of the pulse is $\Omega'=\sqrt{\Delta\omega^2+\Omega^2}$, resulting in additional rotation angle errors.}
\label{fig:Bloch}
\end{figure}

Computation of the coherence amplitudes for \CPMG\ and \UDD\ pulse spacings shows negligible differences in the coherence amplitudes as a result of the different pulse spacing for typical values of the echo time $T_E$ and $T_1$ and $T_2$ relaxation times.  However, Fig.~\ref{fig:pathways} shows that the difference between \CPMG\ and \UDD\ lies in the timings at which the echoes occur.  The many possible coherence pathways give rise to many echoes for both sequences.  However, for \CPMG\ pulse spacing many of these echoes coincide, leaving only a limited number of well-separated echoes.  This endows the \CPMG\ sequence with significant robustness with regard to pulse imperfections --- even if other pathways are populated, most of these are refocused to form echoes concomitant with the target echo or well before or after the target echo.  The latter are likely to be almost entirely dephased during the measurement of the target echo and thus irrelevant.

\UDD\ pulse spacing, on the other hand, results in a multitude of possible echoes, several of which occurring close to but not concomitant with the target echo.  Hence, if some of these other pathways are populated, we are likely to see interference of multiple echoes around the target time, resulting in complicated measurement signals.  At the least the different coherence levels of the pathways, if excited, will reduce the overall coherence of the ensemble and thus the magnitude of the (transverse magnetization) readout signal.  Thus, \UDD\ is more sensitive than \CPMG\ to imperfect excitation and refocusing pulses unless the crusher gradients are large enough to completely dephase all unwanted excitation pathways.  As the differences in the coherence amplitudes of the various pathways for \UDD\ and \CPMG\ are marginal (at least for $n=4$), if unwanted pathways are completely dephased the differences between \UDD\ and \CPMG\ should decrease.  Dephasing of the population in these pathways will reduce the magnetization signal in both cases.

Interference of echo pathways is a likely explanation why \UDD\ appears to enhance inhomogeneities in tissue.  Such inhomogeneities tend to result in slight variations in the effective amplitudes of the RF pulses, resulting in pulse angle errors that lead to population of non-target pathways and echo interference effects.  Local magnetic susceptibility effects, for example, change the local $B$-field by a factor of $(1+\xi)$, $\xi$ being the susceptibility.  As $\omega \propto B_0$, local susceptibility effects lead to resonance offsets in a voxel $\propto \xi B_0$ and result in rotation axis errors as shown in Fig.~\ref{fig:Bloch}.  Thus \UDD\ can enhance local susceptibility effects.  However, it will also enhance inhomogeneities resulting from other sources, such as intrinsic $B_0$ or $B_1$  inhomogeneities, for example.  

\begin{figure*}
\center\includegraphics[width=0.7\textwidth]{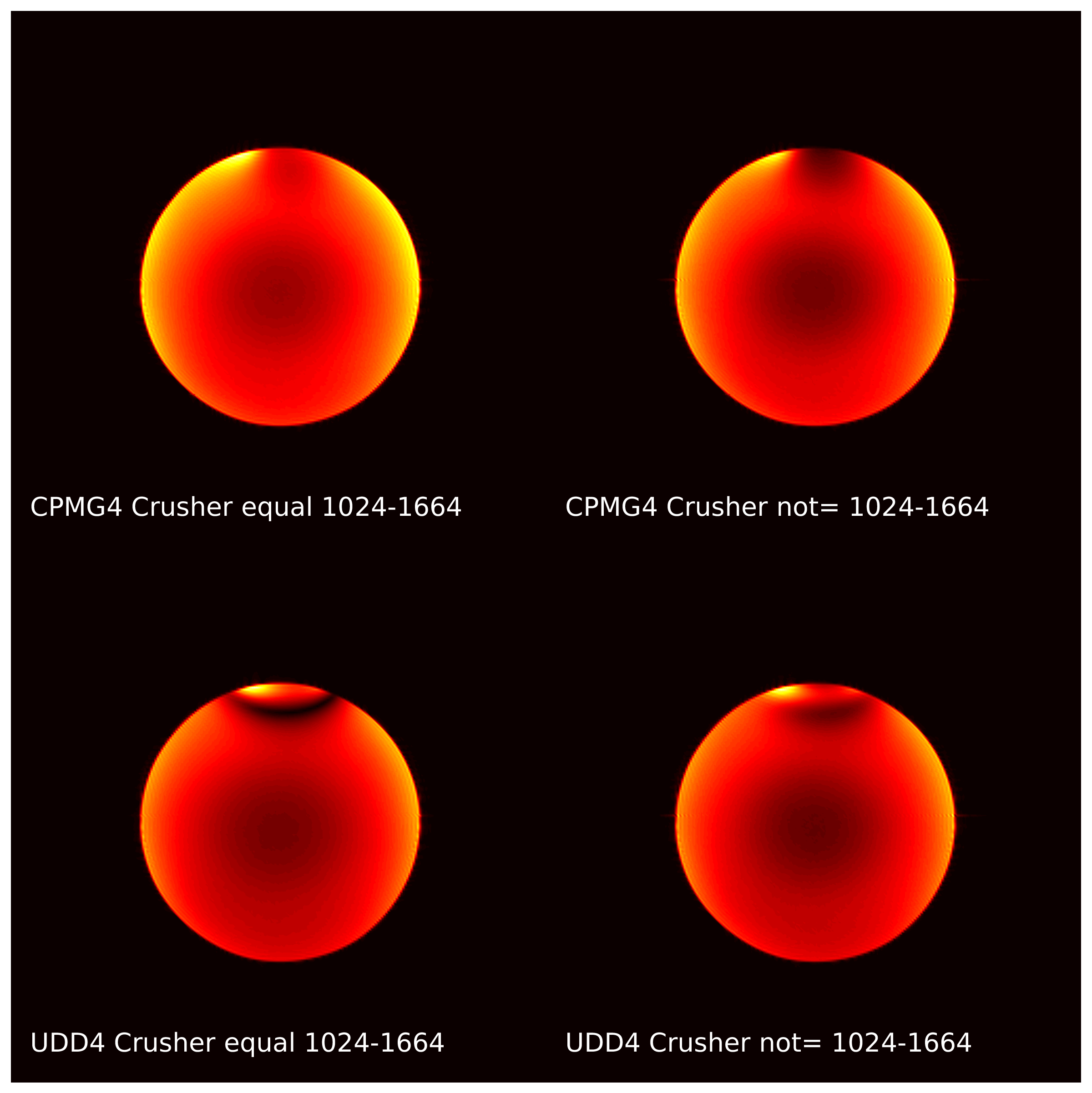}
\caption{\CPMG-4 and \UDD-4 images of a slice of a uniform water phantom with a small aluminum block affixed on top for different choices of crushers.}
\label{fig:susceptibility}
\end{figure*}

The echo interference effect due to local susceptibility variations can be seen in the images of a slice through a cylindrical water phantom with a small aluminum block (not visible) affixed on top in Fig.~\ref{fig:susceptibility}.  The top left \CPMG$_4$ image with uniform crusher gradients shows almost no evidence of the aluminum block as the stimulated echo coherences created in the non-target pathways are refocused and create echoes concomitant with the target echo.  When the crusher gradient moments are made unequal to dephase the population in the non-target pathways a dark spot becomes visible in the vicinity of the aluminum block in the top right \CPMG$_4$ image due to reduced population in the target pathway.  For \UDD$_4$, on the other hand, the bottom left image, acquired with uniform crusher gradients, shows a clear interference fringe, which is significantly \emph{reduced} when non-equal crushers are applied (bottom right).  The fact that the fringe is reduced but still visible is most likely due to the applied crusher gradients being insufficiently strong to completely dephase the non-target pathways.

Chemical shifts can have a similar effect.  Assuming the excitation and refocusing pulses are calibrated to produce perfect $90^\circ$ and $180^\circ$ rotations for water protons then chemical shift offsets, \textit{e.g.}, for oil or fat, result in imperfect rotations that spread the population over many coherence pathways producing effects similar to $B_0$ or $B_1$ inhomogeneity.

\begin{figure*}
\subfloat{\includegraphics[width=0.52\textwidth]{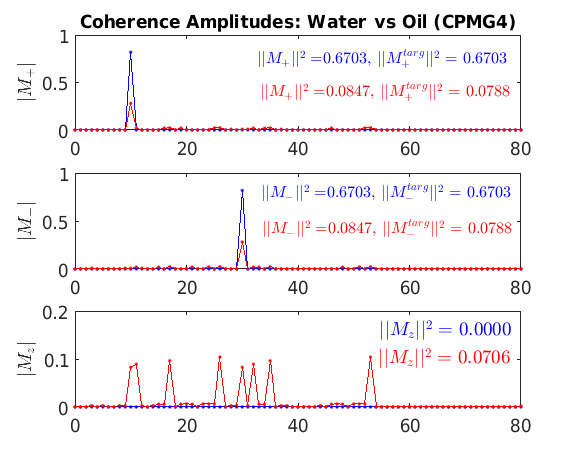} \hspace{-3ex}}
\subfloat{\includegraphics[width=0.52\textwidth]{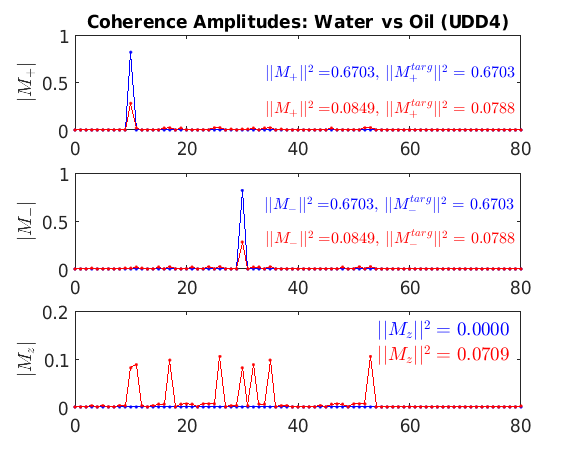}\hspace{-3ex}} 
\caption{Coherence amplitudes for \CPMG$_4$ and \UDD$_4$ with $T_E=40$ for water ($T_1=\SI{4}{\second}$, $T_2=\SI{400}{\milli\second}$) and oil ($T_1=\SI{250}{\milli\second}$, $T_2=\SI{70}{ms}$, chemical shift $\SI{-440}{\hertz}$), assuming pulses perfectly  calibrated for water ($\pi/2$-pulse duration $t_a=\SI{1024}{\micro\second}$, $\pi$-pulse duration $t_b=\SI{1280}{\micro\second}$).} \label{fig:amp_oil}
\end{figure*}

\begin{example}[Water-Oil Chemical Shift Effect.]
The chemical shift for oil relative to water at \SI{3}{\tesla} is \SI{-440}{\hertz}.  If we apply a \SI{1024}{\micro\second} $90^\circ$-pulse at the resonance frequency of water then the nominal Rabi frequency is $\Omega=\pi/(2\times \SI{1.024}{ms}) \approx \SI{1.53}{\kilo\hertz}$.  Hence, $\Delta\omega =-0.287\times\Omega$, the actual Rabi frequency $\Omega'=1.05\times\Omega$ and azimuthal angle of the rotation axis $\theta \approx 1.18\times 90^\circ$, \textit{i.e.}, the rotation angle error is 5\% and rotation axis error is 18\% for the excitation pulse. Similarly, for a \SI{1280}{\micro\second} nominal $180^\circ$-pulse we have $\Omega \approx \SI{2.45}{\kilo\hertz}$ and $\theta=1.11\times 90^\circ$ and $\Omega' = 1.016\times\Omega$,
\textit{i.e.}, a rotation axis error of 11\% and angle error of about 1\%.
\end{example}

The effect of these pulse errors on the coherence amplitudes is shown in Fig.~\ref{fig:amp_oil}.  The amplitudes for oil are overall lower due to much greater relaxation.  The off-resonant pulses convert a significant amount of magnetization to longitudinal magnetization, and of the remaining transverse magnetization, the population of the target pathway is only 93\%.  The differences in the coherence amplitudes between \UDD\ and \CPMG\ are negligible but as \CPMG\ refocuses the populations of the main alternative pathways at $T_E$ along with the main pathway, while \UDD\ does not, \CPMG\ will result in the appearance of better refocusing, especially for uniform or no crusher gradients.

Finally, we observe that the resulting longitudinal coherences created by imperfect pulses decay slowly, having decayed only by $e^{-(T_R-T_E)/T_1}$ at the beginning of the next scan, even in a single echo scan.  Unless $T_R-T_E \gg T_1$ this may result in complex features and characteristics commonly seen with short-$T_R$ steady-state free precession sequences.

\new{To better understand the performance of dynamic decoupling sequences such as \UDD\ in the presence of various experimental imperfections, and optimize the sequence parameters, the coherence pathway analysis could be extended to include full Bloch equation simulations incorporating experimental slice profiles for the excitation and refocusing pulses to calculate the predicted echo signal~\cite{Lebel2010, Petrovic2015}.}

\section{Conclusion}

Despite suggestions in previous work that \UDD\ may provide better refocusing of tissue than the conventional and widely used \CPMG\ sequence~\cite{Warren2009,Warren2013}, our relaxometry results for carefully calibrated experiments using tissue-mimicking gel phantoms suggest that the conventional \CPMG\ sequence with regular spacing of the refocusing pulses provides better refocusing of tissue and greater robustness than \UDD.

These results are supported by theoretical considerations and simulations.  \UDD\ has been shown to provide superior decoupling for a two-level system coupled to a bath characterized by a power spectrum of the form $P(f) \propto f$ with a high-frequency cutoff~\cite{Uhrig2007}.  Our simulations suggest that \UDD\ still performs marginally better than \CPMG\ if the resonance frequency offset is characterized by a blue noise amplitude spectrum.  However, in the context of MRI, no refocusing is preferable to either \CPMG\ or \UDD\ refocusing in the presence of white or blue noise as the effects of resonance offset fluctuations $\Delta\omega(t)$ with frequencies much greater than the inverse echo time, $1/T_E$, tend to average and refocusing interferes with this averaging effect.  The simulations show that refocusing is beneficial in the presence of pink noise, characterized by a $1/f$ amplitude spectrum and dominated by low-frequency fluctuations.  Both theoretical and empirical observations suggest that biological systems such as tissue are generally subject to $1/f$ noise.  Our simulations show that \CPMG\ outperforms \UDD\ even for ideal instantaneous refocusing pulses for pink noise, consistent with our empirical observations.

Coherence pathway analysis furthermore suggests that regular pulse spacing such as \CPMG\ benefits from robustness with regard to imperfect excitation and refocusing pulses.  Imperfect pulses lead to the accumulation of population in a multitude of coherence pathways and the formation of a multitude of echoes in addition to the main echo at time $T_E$.  For the regular pulse spacing of \CPMG\ many of these additional echoes coincide with the main echo.  This simultaneous refocusing of alternative echo pathways endows \CPMG\ with special robustness in the presence of imperfect pulses.  For non-uniform spacing of the refocusing pulses, as in \UDD, additional echoes occur at different times.  For \UDD$_4$ spacing, for example, there are three additional echoes offset from the main echo by less than 10\%, which can interfere, producing interference patterns in the image.  This effect renders sequences such as \UDD\ more susceptible to pulse imperfections and both $B_0$ and \RF\ inhomogeneity.

While the sensitivity to $B_0$ inhomogeneity is problematic in the presence of inhomogeneity caused by technical imperfections of the equipment,  it also renders \UDD\ more sensitive to resonance offsets due to chemical shifts --- which may explain some of the differences observed in the refocusing of oil and water --- as well as sensitivity to small variations in tissue susceptibility -- which may explain the apparent enhancement of inhomogeneities in tissue observed in \cite{Warren2009} --- and suggests that decoupling sequences with non-equal pulse spacing may be useful for certain imaging tasks.  Given that \UDD\ spacing is less effective than \CPMG\ for $1/f$ noise, however, we conjecture that decoupling sequences with non-equal spacing different from \UDD\ spacing exist that outperform \UDD\ and \CPMG\ in refocusing tissue.  The design of such sequences requires further experimental characterization of noise spectra for different types of healthy and diseased tissue.

Another area of future work suggested by our results are dynamic decoupling sequences optimized for realistic \RF\ pulses.  Although it has been observed in \cite{Uhrig2010} that in principle, the effect of finite pulse durations can be mitigated by using optimized, shaped pulses, in MRI the need for slice-selective pulses imposes constraints on the duration of the pulses and optimization of pulse shapes.  This limits the number of decoupling pulses that can be applied and suggests a need for more robust, simultaneously time and slice-profile optimized \RF\ pulses.  Furthermore, much greater static magnetic and \RF\ field inhomogeneity in large-bore \SI{1.5}{T} or \SI{3}{T} clinical MR systems compared to high-field NMR spectrometers, make robustness of dynamic decoupling sequences to such imperfections critical for applications in MRI.

\section*{References}

\bibliographystyle{iopart-num}
\bibliography{../references/UDD.bib}
  
\appendix

\section{$T_2$ fits for \CPMG$_n$ and \UDD$_n$ for all samples}
\label{app:T2}

\begin{figure*}[h]
  \includegraphics[width=0.49\textwidth]{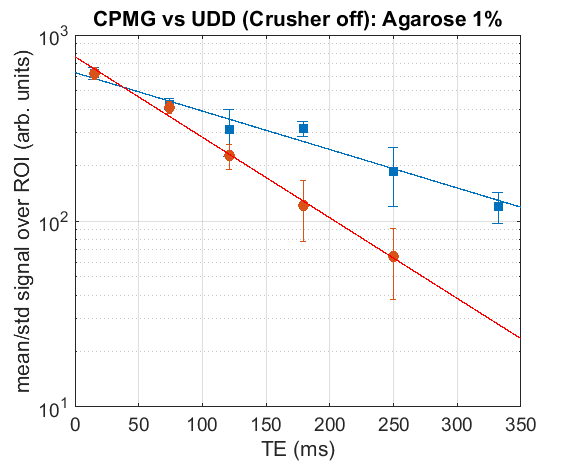} \hfill
  \includegraphics[width=0.49\textwidth]{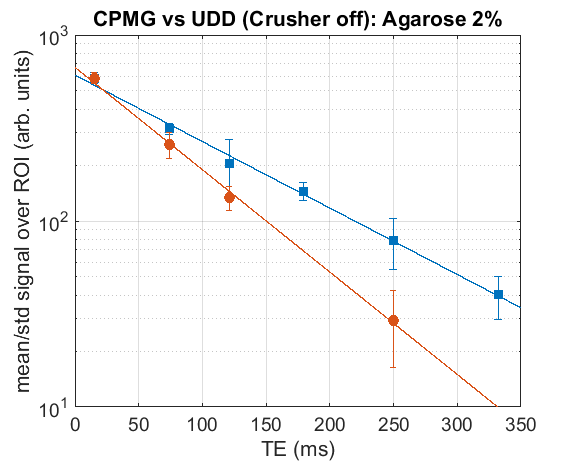}\\
  \includegraphics[width=0.49\textwidth]{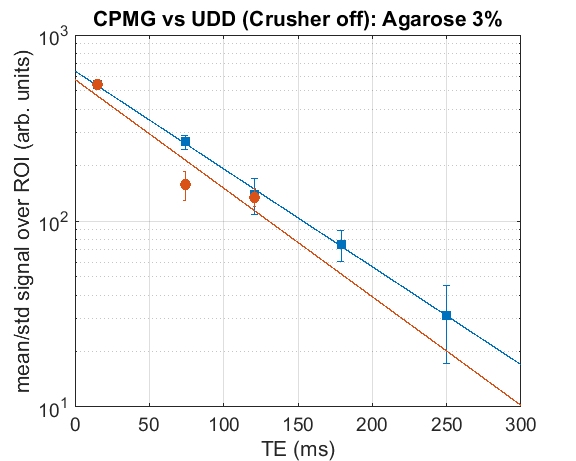}\hfill
  \includegraphics[width=0.49\textwidth]{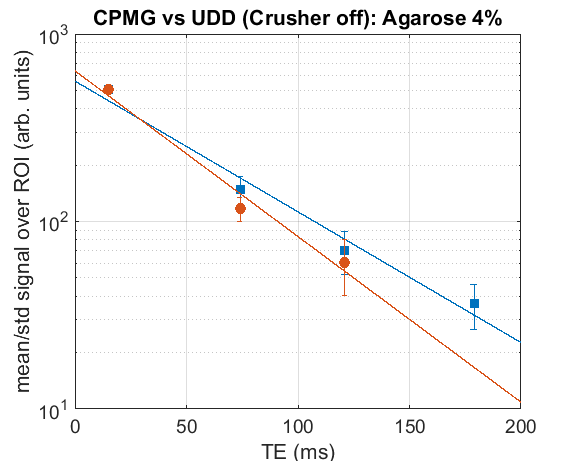}\\
  \includegraphics[width=0.49\textwidth]{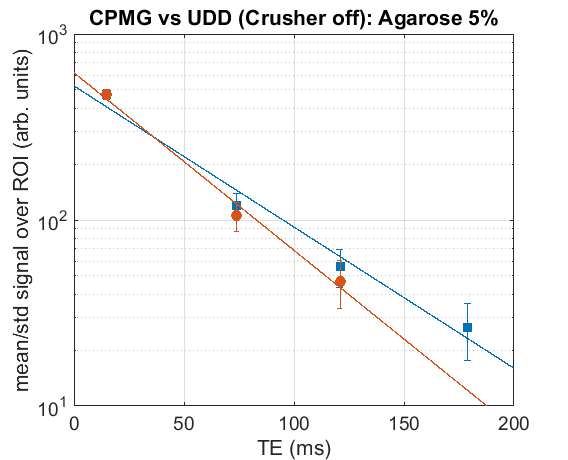}\hfill
\caption{Agarose gel $T_2$ fits for \CPMG$_n$ (blue) and \UDD$_n$ (red) and $R_2$ vs concentration plot without crusher gradients.} \label{fig:T2-fits1}
\end{figure*}
\begin{figure*}
  \includegraphics[width=0.49\textwidth]{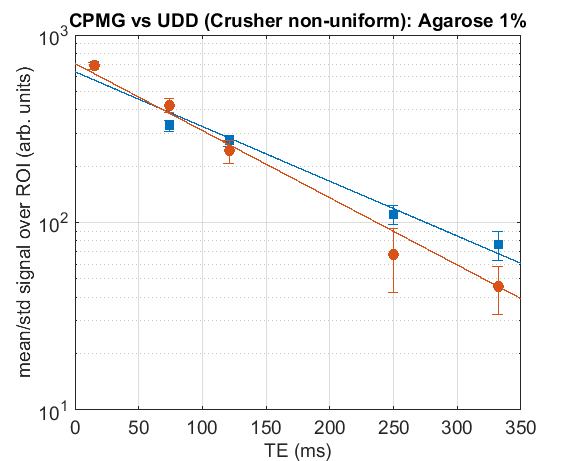}\hfill 
  \includegraphics[width=0.49\textwidth]{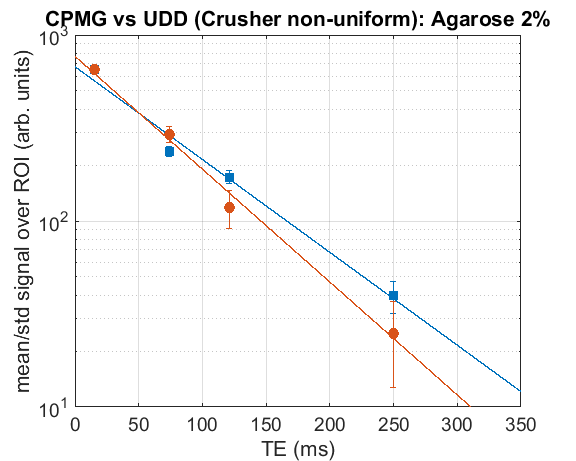}\\
  \includegraphics[width=0.49\textwidth]{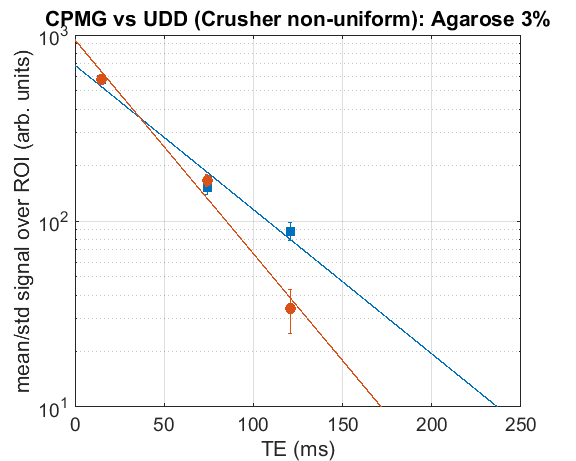}\hfill
  \includegraphics[width=0.49\textwidth]{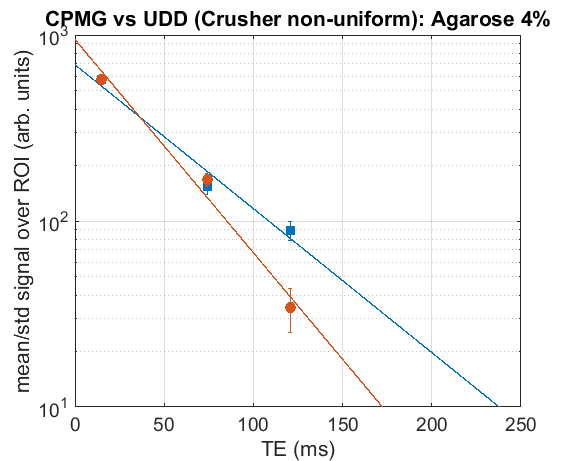}\\
  \includegraphics[width=0.49\textwidth]{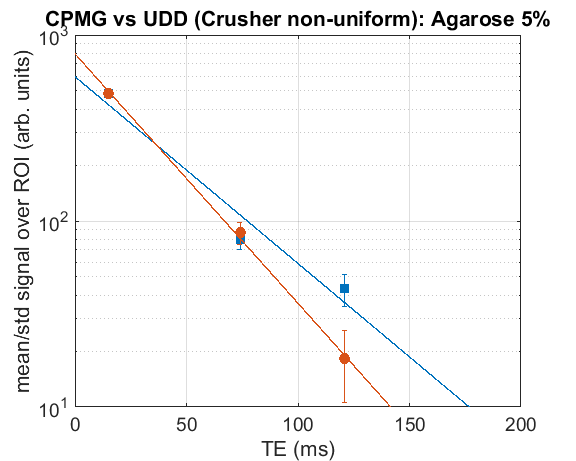}\hfill
\caption{Agarose gel $T_2$ fits for \CPMG$_n$ (blue) and \UDD$_n$ (red) and $R_2$ vs concentration plot with nonuniform crusher gradients.} \label{fig:T2-fits2}
\end{figure*}

\begin{figure*}
  \includegraphics[width=0.49\textwidth]{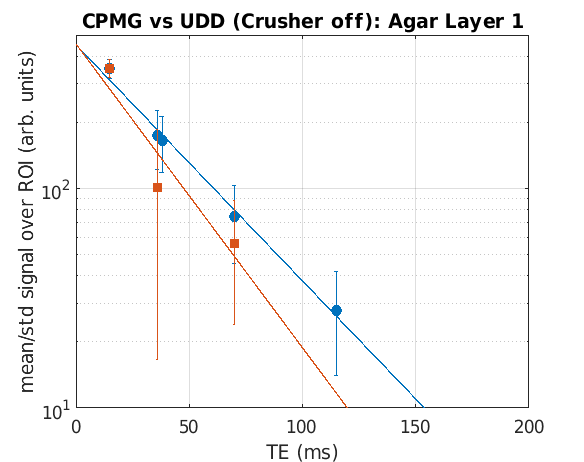} \hfill
  \includegraphics[width=0.49\textwidth]{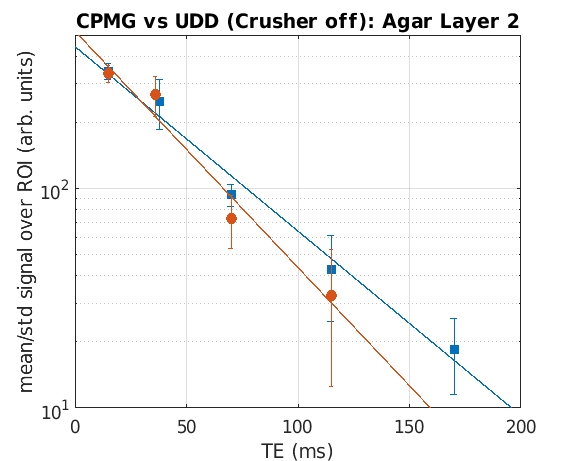}\\ 
  \includegraphics[width=0.49\textwidth]{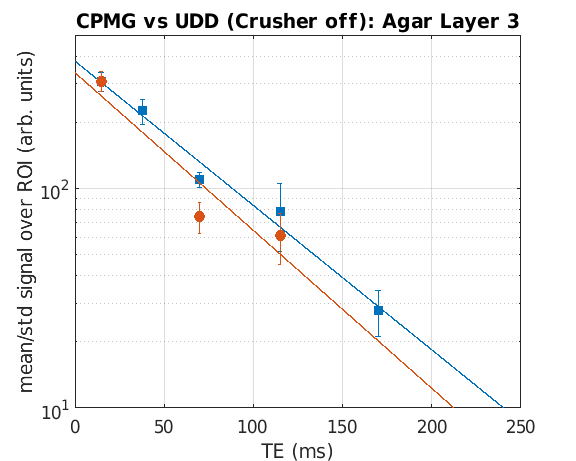}\hfill
  \includegraphics[width=0.49\textwidth]{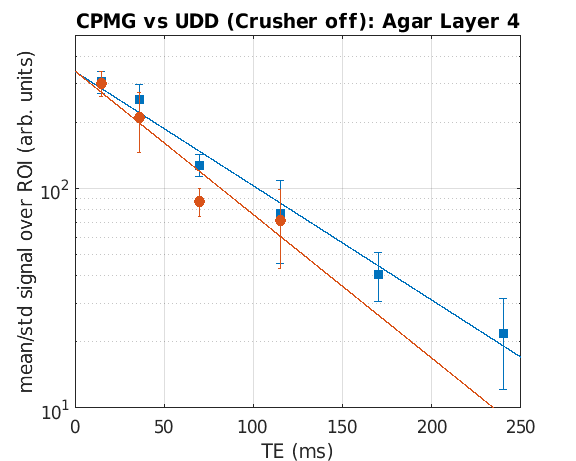}\\
  \includegraphics[width=0.49\textwidth]{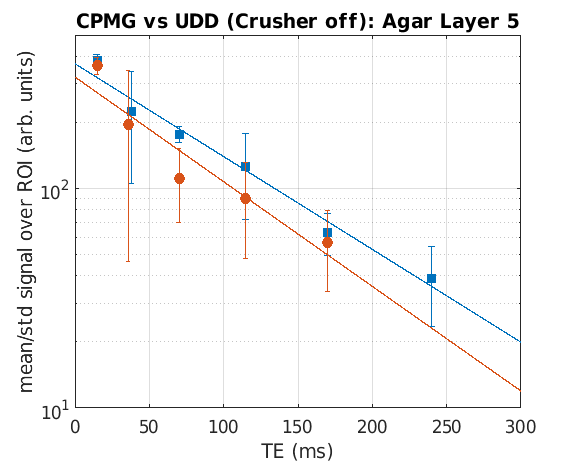}\hfill
  \includegraphics[width=0.49\textwidth]{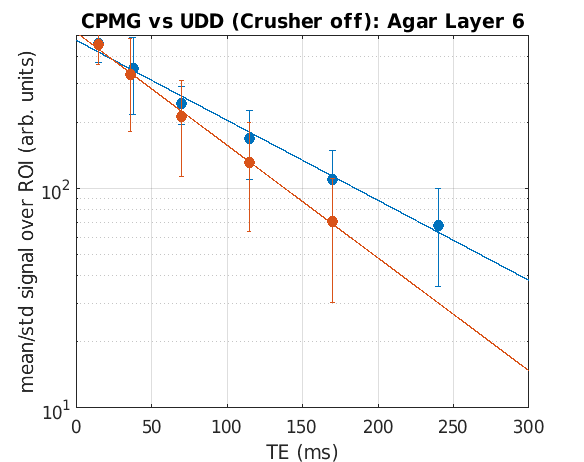} 
\caption{Agar layer gel $T_2$ fits for \CPMG$_n$ (blue) and \UDD$_n$ (red) with uniform crusher gradients.} \label{fig:T2-fits3}
\end{figure*}

\begin{figure*}
  \includegraphics[width=0.49\textwidth]{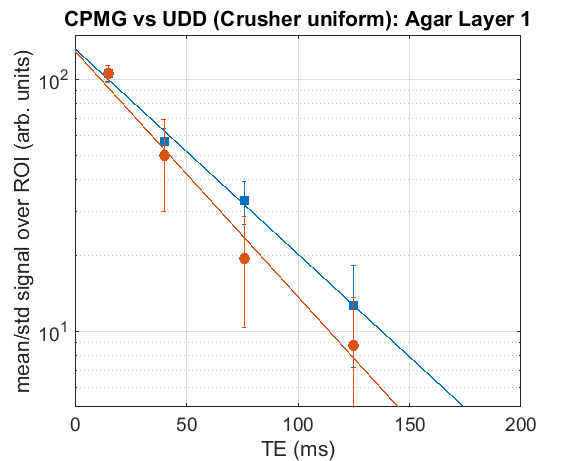} \hfill
  \includegraphics[width=0.49\textwidth]{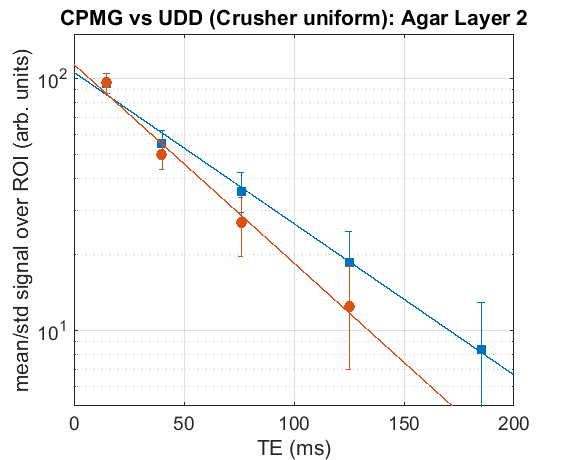}\\ 
  \includegraphics[width=0.49\textwidth]{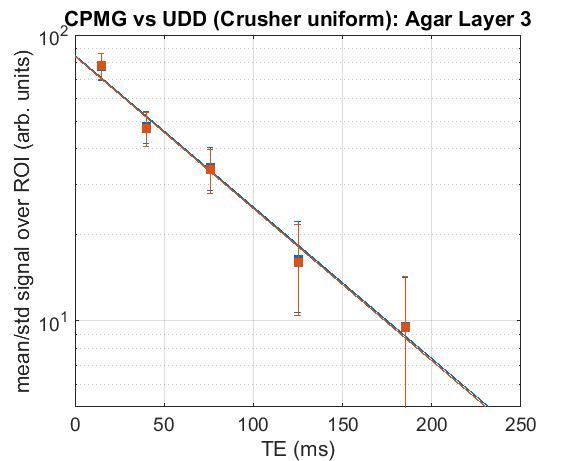}\hfill
  \includegraphics[width=0.49\textwidth]{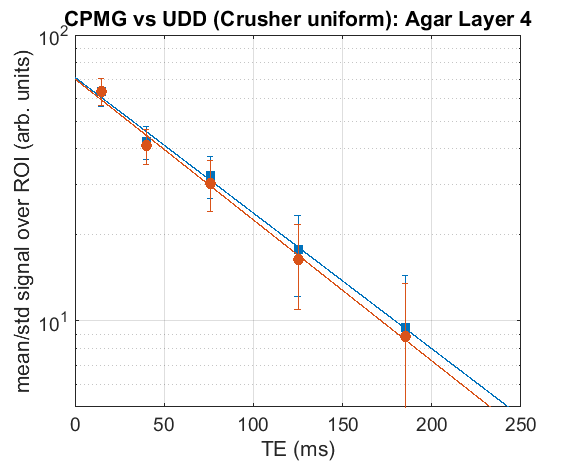}\\
  \includegraphics[width=0.49\textwidth]{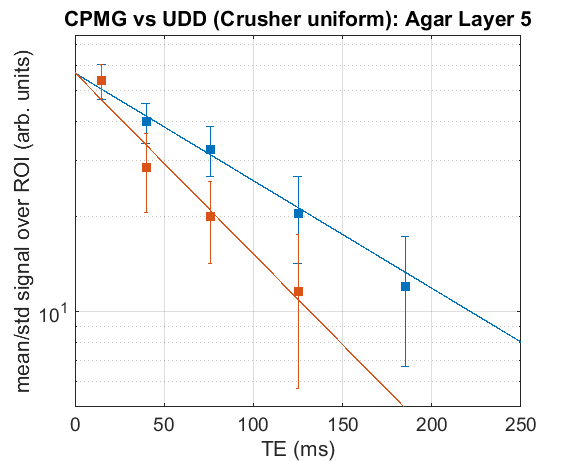}\hfill
  \includegraphics[width=0.49\textwidth]{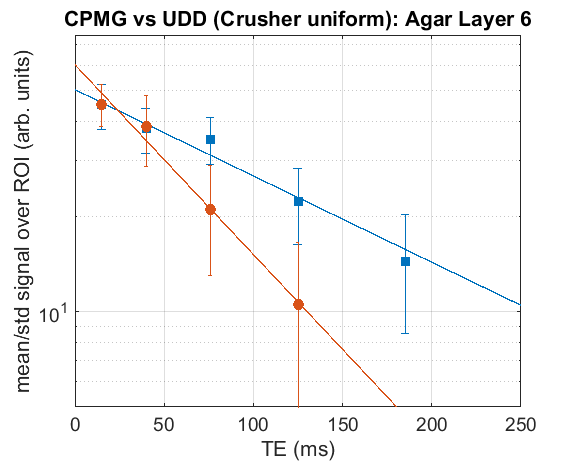} 
\caption{Agar layer gel $T_2$ fits for \CPMG$_n$ (blue) and \UDD$_n$ (red) with uniform crusher gradients.} \label{fig:T2-fits4}
\end{figure*}

\end{document}